\RequirePackage{fix-cm}
\documentclass[smallextended]{svjour3}       
\smartqed  

\usepackage{natbib}
\usepackage{booktabs}
\usepackage{graphicx}
\usepackage{amsmath}
\usepackage{amssymb}
\usepackage{paralist}
\usepackage[font=small,labelfont=bf]{caption}
\usepackage{multirow}
\usepackage{xspace}
\usepackage{color}
\usepackage{xcolor}
\usepackage{ifthen}
\usepackage{url}
\usepackage{fancybox}
\usepackage{enumitem}
\usepackage{listings}
\usepackage{balance}
\usepackage{ifthen}
\usepackage[tight,footnotesize]{subfigure}
\usepackage{flushend}
\usepackage{tablefootnote}
\newenvironment{hassanbox}%
{\begin{center}\vspace{0mm}\noindent\begin{Sbox}\begin{minipage}{0.95\columnwidth}}%
{\end{minipage}\end{Sbox}\fbox{\TheSbox}\end{center}\vspace{0mm}}

\usepackage{color}
\usepackage{ifthen}
\newboolean{showcomments}
\setboolean{showcomments}{true} 
\ifthenelse{\boolean{showcomments}}
  {

        \newcommand{\dmg}[1]{{\color{blue}\emph{dmg says: #1}}\xspace}
        \newcommand{\todo}[1]{\textcolor{red}{{\sc #1}}}
        \newcommand{\internalnote}[1]{\marginpar{\scriptsize note: #1}}
        
		\newcommand{\remarks}[1]{\color{red}[#1]\color{black}}

		\newcommand{\del}[1]{\textcolor{red}{\sout{#1}}} 
  }
  {
        \newcommand{\dmg}[1]{}
        \newcommand{\todo}[1]{}
        \newcommand{\internalnote}[1]{}
		\newcommand{\remarks}[1]{}

		\newcommand{\del}[1]{} 
  }

\usepackage{color}
\definecolor{gray}{rgb}{0.4,0.4,0.4}
\definecolor{darkblue}{rgb}{0.0,0.0,0.6}
\definecolor{cyan}{rgb}{0.0,0.6,0.6}
\definecolor{backcolour}{rgb}{0.95,0.95,0.92}

\lstdefinelanguage{XML}
{
	backgroundcolor=\color{backcolour},
	morestring=[b]",
	morestring=[s]{>}{<},
	morecomment=[s]{<?}{?>},
	stringstyle=\color{black},
	identifierstyle=\color{darkblue},
	keywordstyle=\color{cyan},
	morekeywords={dependency,version,dependencies}
}

\lstdefinelanguage{email}
{
	backgroundcolor=\color{backcolour},
	morestring=[b]",
	morestring=[s]{>}{<},
	morecomment=[s]{<?}{?>},
	stringstyle=\color{black},
	identifierstyle=\color{cyan},
	keywordstyle=\color{cyan},
	morekeywords={}
}
\DeclareGraphicsExtensions{.pdf,.jpeg,.png}

\newcommand{\ie}{i.e.,}

\newcommand{\dmc}{DU}
\newcommand{\dmcfull}{Dependency Update}

\newcommand{\lib}[2]{$\mathcal{L}$(\texttt{#1},\textit{#2)}}
\newcommand{\sys}[2]{$\mathcal{S}$(\texttt{#1},\textit{#2)}}

\newcommand{\RqOne}{\emph{ To what extent are developers updating their library dependencies?}\xspace}

\newcommand{\RqTwo}{\emph{What is the response to important awareness mechanisms such as a new release announcement and a security advisory on library updates?}\xspace}

\newcommand{\RqThree}{\emph{Why are developers non responsive to a security advisory?}\xspace}

\begin{document}

\title{Do Developers Update Their Library Dependencies?}
\subtitle{An Empirical Study on the Impact of Security Advisories on Library Migration}


\author{Raula Gaikovina Kula \and Daniel M. German \and
 Ali Ouni         \and
 Takashi Ishio	\and
        Katsuro Inoue
}


\institute{Raula Gaikovina Kula, Ali Ouni, Takashi Ishio and Katsuro Inoue  \at
		  Osaka University, Japan \at
          \email{\{raula-k, ali, ishio, inoue\}@ist.osaka-u.ac.jp}          
           \and
           Daniel M. German \at
           University of Victoria, Canada \at
           \email{dmg@uvic.ca}          
}

\date{Received: date / Accepted: date}

\maketitle
\begin{abstract}
Third-party library reuse has become common practice in contemporary software development, as it includes several benefits for developers.
Library dependencies are constantly evolving, with newly added features and patches that fix bugs in older versions.
To take full advantage of third-party reuse, developers should always keep up to date with the latest versions of their library dependencies.
In this paper, we investigate the extent of which developers update their library dependencies.
Specifically, we conducted an empirical study on library migration that covers over 4,600 GitHub software projects and 2,700 library dependencies.
Results show that although many of these systems rely heavily on dependencies, 81.5\% of the studied systems still keep their outdated dependencies.
In the case of updating a vulnerable dependency, the study reveals that affected developers are not likely to respond to a security advisory.
Surveying these developers, we find that 69\% of the interviewees claim that they were unaware of their vulnerable dependencies. 
Furthermore, developers are not likely to prioritize library updates, citing it as extra effort and added responsibility.
This study concludes that even though third-party reuse is commonplace, the practice of updating a dependency is not as common for many developers.
                               

\end{abstract}

\section{Introduction}
In contemporary software development, developers often rely on third-party libraries to provide a specific functionality in their applications.
In 2010, Sonatype reported that Maven Central\footnote{\url{http://search.maven.org/}} contained over 260,000 maven libraries\footnote{Link at  \url{http://goo.gl/SV9d68}}.
As of November 2016, this collection of libraries rose to 1,669,639 unique Maven libraries\footnote{statistics accessed Nov-26th-2016 at \url{https://search.maven.org/\#stats}}, which is almost six times more than it was in 2010 and making it one of the largest hosting repositories of OSS libraries.
Libraries aim to save both time and resources and reduce redundancy by taking advantage of existing quality implementations.

Many libraries are in constant evolution, releasing newer versions that fix defects, patch vulnerabilities and enhance features.
In fact, \cite{Lehman:1996} states that software either \textit{`undergoes continual changes or becomes progressively less useful'. }
As software development transitions into the maintenance phase, a developer becomes the maintainer and is faced with the following software maintenance dilemma:
  \textit{`When should I update my current library dependencies?'}  
We define this dilemma of updating libraries as the \textit{library migration} process, which involves movement from a specific library version towards a newer replacement version of the same library, or to a different library altogether.

The decision to migrate a library can range from being rather trivial to extremely difficult.
Typically, a developer evaluates the overall quality of the new release version, taking into account: (i) new features, (ii) compatibility compared to the current version, (iii) popular usage by other systems and (iv) documentation, support and longevity provided by the library.
On the other hand, migration of a vulnerable dependency requires an immediate response from the developer.
It is strongly recommended to immediately migrate a vulnerable dependency, as it exposes the dependent application to malicious attacks.  
In response to these vulnerabilities, emergence of \textit{awareness mechanisms} such as the \texttt{Common Vulnerabilities and Exposures} (CVE)\footnote{\url{http://cve.mitre.org/cve/index.html}} database are designed to raise developer awareness and trigger the migration of a vulnerable dependency.

In this paper, we investigate the extent of how library migration is practiced in the real-world.
Our goals are to investigate (1) whether or not library dependencies are being updated and (2) the level of developer awareness to library migration opportunities.
Specifically, we performed a large-scale empirical study to track library migrations between an application client (defined as a system) and their dependent library provider (defined as a library).
The study encompasses 4,659 projects, 8 case studies and a developer survey to draw the following conclusions:

\noindent \textit{(1) Library Migration in Practice:}
Although systems depend heavily on libraries, findings show that many of these systems rarely update their library dependencies.
Developers are less likely to migrate their library dependencies, with up to 81.5\% of systems keeping outdated dependencies.

\noindent \textit{(2) Developer Responsiveness to Awareness Mechanisms:}
Our findings indicates patterns of either consistent migration or a lack of library migration. 
We find many cases where developers prefer an older and popular dependency over a newer replacement.  
Importantly, the study depicts developers as being non responsive to a security advisory.
In a follow-up survey of affected developers, 69\% of the interviewees claim that they were unaware of the vulnerability and who then promptly migrated away from that vulnerable dependency.
Furthermore, developers cite (i) a lack of awareness in regard to library migration opportunities, (ii) impact and priority of the dependency, and (iii) the assigned roles and responsibilities as deciding factors on whether or not they should migrate a library dependency.

\sloppypar{
Our main contributions are three-fold.
Our first contribution is a study on library migration pertaining to developer responsiveness to existing awareness mechanisms (\ie~security advisory).
Our second contribution is the modeling of library migration from system and library dimensions, with different metrics and visualizations such as the Library Migration Plot (LMP).
Finally, we make available our dataset of 852,322 library dependency migrations.
All our tools and data are publicly available from the paper's replication package at \url{https://raux.github.io/Impact-of-Security-Advisories-on-Library-Migrations/}.
}



\subsection{Paper Organization}
The rest of the paper is organized as follows. Section \ref{sec:back} describes the basic concepts of library migrations and awareness mechanisms.
Section \ref{sec:RQ} motivates our research questions, while Section \ref{sec:method} describes our research methods to address them.
The results and case studies of the empirical study are presented in Section \ref{sec:prac} and Section \ref{sec:LMT}.
We then discuss implications of our results and the validity threats in Section \ref{sec:dis}, with Section \ref{sec:related} surveying related works.
Finally, Section \ref{sec:conclude} concludes our paper.

\section{Basic Concepts \& Definitions}
\label{sec:back}                      

In this section, we introduce the library migration process and the related terminologies that will be used in the paper.
Building on our previous work of trusting the latest versions of libraries \citep{KulaSANER2014} and visualizing the evolution of libraries \citep{2014VISSOFTKula}, this paper is concerned with empirically tracking library migration and understanding the awareness mechanisms that trigger the migration process. 
We first present the library migration process in Section \ref{sec:mp}. 
Then later in Section \ref{sec:am}, we introduce two common awareness mechanisms that are designed to trigger a library migration.

\subsection{The Library Migration Process}
\label{sec:mp}
We identify these three generic steps performed by a developer during the library migration process:

\begin{itemize}
	\item \underline{\textit{ Step 1: Awareness of a Library Migration Opportunity}}.
	Step 1 is triggered when a developer becomes aware of an opportunity to migrate a specific dependency.
	The awareness mechanism may be in the form of either a new release announcement or a security advisory by authors of the library.
	In order for a successful migration, a developer must also identify a suitable replacement for the current dependency.
	In the case of a vulnerable dependency, a developer must identify a safe (patched) library version as a viable replacement candidate for the migration.

	\item \underline{\textit{ Step 2: Migration Effort to Facilitate the Replacement Dependency}}.
	Step 2 involves the efforts of a developer to ensure that the replacement dependency is successfully integrated into the system.
	Specifically, we define this \textit{migration effort} as the amount of work and testing needed to facilitate the replacement dependency.
	This step may involve writing additional integration code and testing to make sure that the replacement library does not break current functionality, or affect other dependencies that co-exist within the system.

	\item\underline{\textit{Step 3: Performing the Library Migration}}.
	Step 3 ends the library migration process. 
	Once the migration effort in Step 2 is completed, the prior dependency is then abandoned, with the replacement library adopted by the system.
\end{itemize}


\subsection{Library Migration Awareness Mechanisms}
\label{sec:am}
To trigger the library migration process, developers must first become aware of the necessity to migrate a dependency.
In this section, we discuss the two most common types of awareness mechanisms that include (1) a new version release announcement and (2) a security advisory.

 \begin{figure}
 	\centering
 	\includegraphics[width=.9\textwidth]{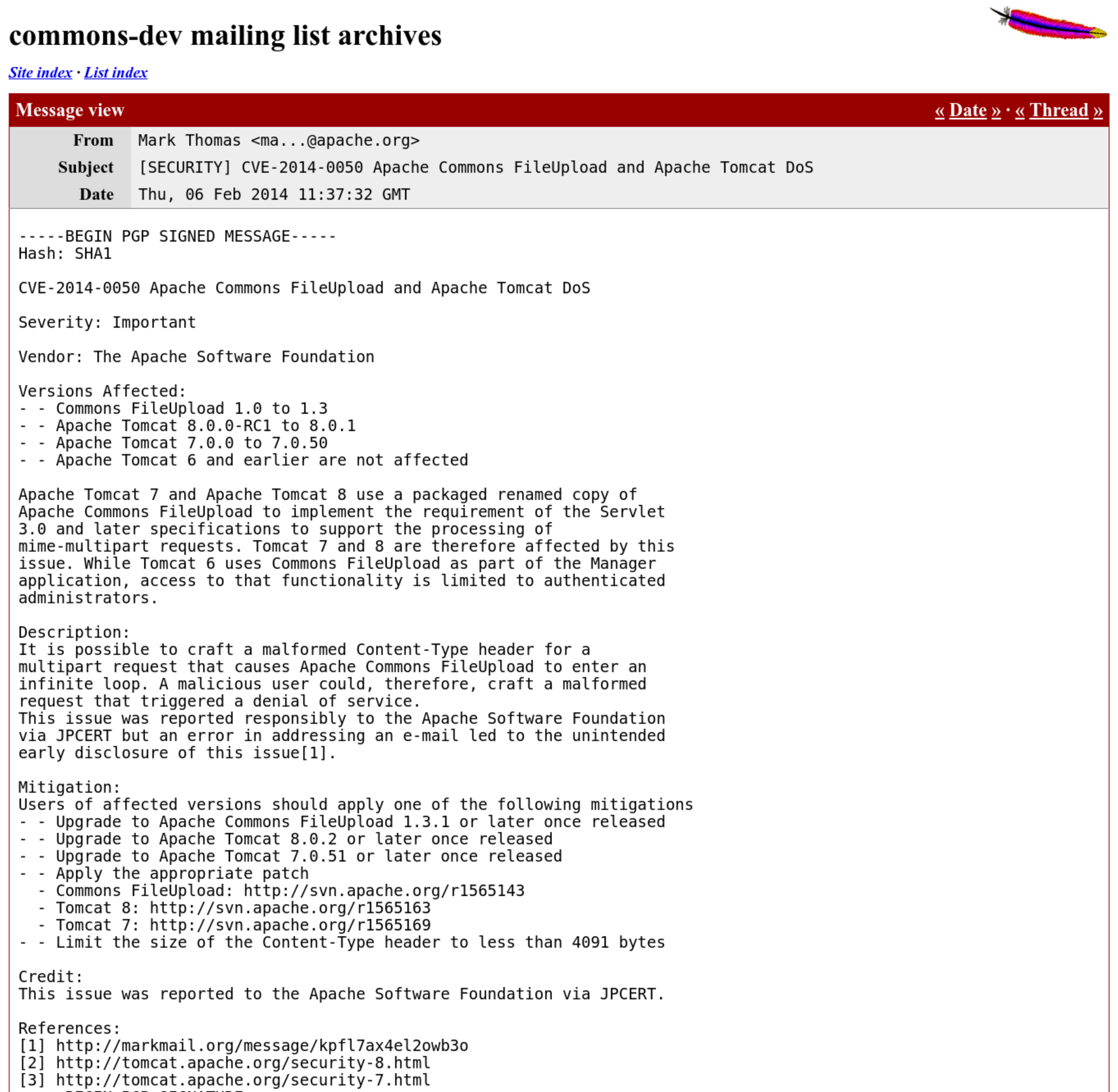}
 	\caption{Example of a security advisory related to \texttt{CVE-2014-0050} that was posted in the Apache common developers mailing list.}
 	\label{fig:securityAdv}
 \end{figure}

\paragraph{\textbf{(1) A New Release Announcement:}}
The traditional method to raise awareness of a new release is through an announcement from the official homepage of the library.
Documentation such as the developer change logs are useful guides to estimate the migration effort needed to perform a successful migration.
In detail, we can infer the migration effort required from the following two sources:
\begin{enumerate}[label=(\roman*)]
	\item \textit{Change logs of releases} - New releases may be caused by newer versions that support the state-of-the-art environments (\ie~support for the Java Development Kit (JDK)). 
	Specific to the library, the change logs detail API changes between releases\footnote{Application Programming Interface (API) changes will result in more migration effort for developers}, new features and fixes to bugs in the prior versions.
	\item \textit{Semantic versioning of releases} - The semantic versioning naming convention\footnote{\url{http://semver.org/}} hints the migration effort needed to perform the migration. For instance, a major released version may require more migration effort than a minor released version of that library.
\end{enumerate}

\paragraph{\textbf{ (2) A Security Advisory}:}
A security advisory is an official public announcement of a verified vulnerable library dependency.
Security advisories are circulated through various mail forums, special mailing lists and security forums with the key objective of raising developer awareness to these vulnerabilities.
Figure \ref{fig:securityAdv} is an example of a mail announcement of the \texttt{CVE-2014-0050} vulnerability sent to Apache Open Source developers and maintainers. 
Vendors and researchers keep track of each vulnerability through a tagged CVE Identifier (\ie~\texttt{CVE-xxx-xxxx}).  
Generally, the advisory contains the following information: (i) a description of the vulnerability, (ii) a list of affected dependencies and (iii) a set of mitigation steps that  usually includes a viable (patched) replacement dependency.

In order to understand the required library migration effort,  we first need to understand the role played by a security advisory in the life-cycle of a vulnerability.
As defined by CVE, a vulnerability undergoes the following four phases:
\begin{enumerate}[label=(\roman*)]
\item \textit{Threat detection} - this is the phase where the vulnerability threat is first discovered by security analysts.
\item \textit{CVE assessment} -  this is the phase where the threat is assessed and assigned a rating by the CVE.
\item \textit{Security advisory} - this is the phase where the threat is publicly disclosed to awareness mechanisms such as the US National Vulnerability Database (NVD)\footnote{\url{https://web.nvd.nist.gov/}} to gain the attention of maintainers and developers.
\item \textit{Patch release} - this is the phase where the library developers provide mitigation options, such as a replacement dependency to patch the threat.
\end{enumerate}
Once a viable replacement dependency (\ie~patch release) becomes available, developers can proceed to complete the library migration process.
There exist cases where the vulnerability life-cycle is not synchronized with the migration process.
For instance, a viable replacement dependency may become available before the security advisory. 
In this case, a developer may migrate their vulnerable dependency before the security advisory is disclosed to the general public.

\section{Research Questions}
\label{sec:RQ}
Our motivation stems from reports of outdated and vulnerable libraries being widespread in the software industry.
In 2014, Heartbleed\footnote{\url{https://web.nvd.nist.gov/view/vuln/detail?vulnId=CVE-2014-0160}}, Poodle\footnote{\url{ https://web.nvd.nist.gov/view/vuln/detail?vulnId=CVE-2014-3566}},  Shellshock\footnote{\url{https://web.nvd.nist.gov/view/vuln/detail?vulnId=CVE-2014-6271}}, --all high profile library vulnerabilities were found to have affected a significant portion of the software industry.
In that same year, Sonatype determined that over 6\% of the download requests from the Maven Central repository were for component versions that included known vulnerabilities.
The company reported that in review of over 1,500 applications, each of them had an average of 24 severe or critical flaws inherited from their components\footnote{report published January 02, 2015 at \url{http://goo.gl/i8J1Zq}}.

The goals of our study is to investigate (1) whether or not dependencies are being updated and (2) the level of developer awareness to dependency migration opportunities.
To do so, we design three research questions that involves a rigorous empirical study and follow-up survey on reasons why developers did not update their library dependencies.
Hence, we first formulate (RQ1) to investigate library migration in practice:
\paragraph{Library Migration in Practice.}

\begin{itemize}
	\item \textbf{(RQ1)} \RqOne
	Prior studies have shown that developer responsiveness to library updates is slow and lagging.
	A study by \cite{Robbes:2012} shows how projects from the Smalltalk ecosystem exhibited a slower reaction to Application Programming Interface (API) updates. 
	Similar results were observed for projects developed in the Pharo \citep{hora:2015} and Java \citep{Sawant2016} programming languages. 
	\cite{Bavota:2015} studies how changes in an Application Programming Interface (API) may trigger library migrations within the ecosystem of Apache products.
	These studies are examples of current literature that has analyzed trends of library usage at the API level of abstraction.

	In this work, we would like to better understand (i) the extent to which developers use third-party libraries and (ii) the migration trends of these libraries.
	Therefore, in (RQ1), we define and model library migration as evolving systems and their library dependencies at a higher abstraction than the API level.
\end{itemize}

In this study, we are particularly interested in the effect of awareness mechanisms on maintainers.
Henceforth, (RQ2) and (RQ3) were formulated to investigate how developers respond to current awareness mechanisms:

\paragraph{Developer Responsiveness to Awareness Mechanisms.}

\begin{itemize}
	\item \textbf{(RQ2)} \RqTwo
	To fully utilize the benefits of a library, developer are recommended to make an immediate response to a library migration opportunity. 
	Therefore, in (RQ2) we study maintainer responsiveness to the awareness mechanisms of (i) new releases and (ii) security advisories.

	\item \textbf{(RQ3)} \RqThree
	Studies show that influencing factors such as personal opinions, organizational structure or technical constraints \citep{Bogart:SCGSE15,Plate:ICSME2015} determines whether or not a developer will migrate a dependency.
	In fact, these studies conclude that developers often \textit{`struggle'} with change, citing current awareness mechanisms as being insufficient.
	However, we conjecture that a vulnerable dependency warrants the immediate attention of all project members. 
	Therefore, in (RQ3) we seek developer feedback to understand why developers would not respond to a vulnerable dependency threat.


\end{itemize}

%
%
%

\section{Research Methods}
\label{sec:method}
In this section, we present the research methods used to address each of the three research questions.
Firstly, to answer (RQ1), we conduct an empirical study by mining and reconstructing historic library migrations for a set of real-world projects.
For (RQ2), we analyze case studies of library migrations pertaining to new releases and vulnerable dependency updates.
Finally to answer (RQ3), we interview developers who currently have vulnerable dependencies in their projects.

\subsection {(RQ1) \RqOne}
Our research method to answer the first research question (RQ1) is a vigorous statistical analysis of library migration for real-world projects.
Our method is comprised of three steps: (1) tracking systems and dependency updates, (2) extraction and analysis system and library dependency measures (3) data collection.
The results of (RQ1) are presented in Section \ref{sec:prac}.

\begin{figure}
	\centering
	\includegraphics[width=1\textwidth]{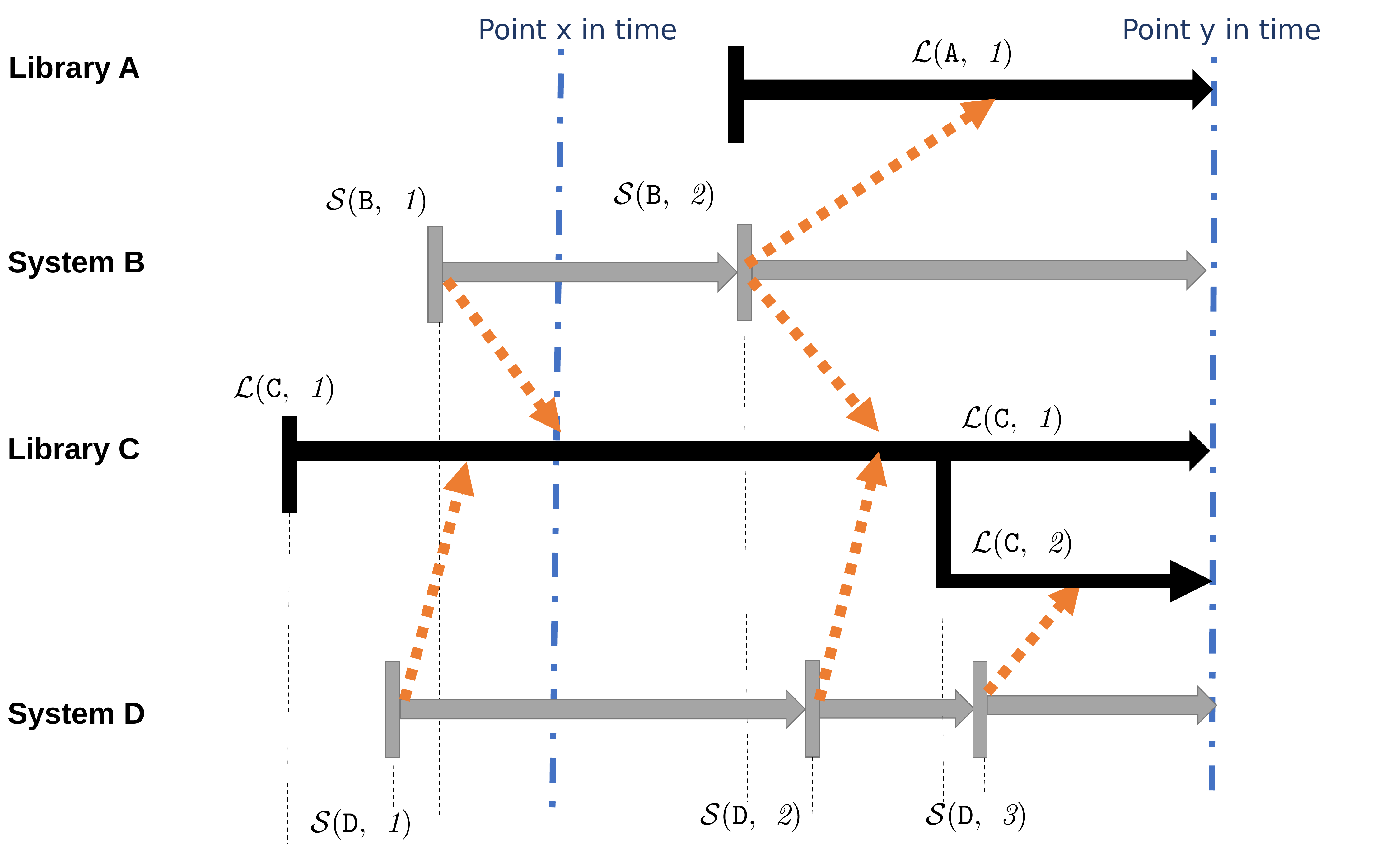}
	\caption{Library migration between systems and libraries. The orange arrow depicts dependency relations between them.}
	\label{fig:migration}
\end{figure}

\paragraph{\textbf{(1) Tracking System and Library Updates}:}
\label{sec:model}
To accurately track dependency migrations, we define a model of system and library dependency relations.
Hence, we formally use the following notations.
We define $\mathcal{S}$ for a system, and $\mathcal{L}$ for a library.
\lib{lib}{v} denotes version $v$ of a library \texttt{lib}, and  \sys{sys}{w} for version $w$ of a system \texttt{sys}.
\label{sec:addDrop}
Adoption of a library version  \lib{lib}{v} by a system version \sys{sys}{w} creates a dependency relation between them.

Figure~\ref{fig:migration} illustrates the notation used to represent the dependency relations between systems and libraries over time.
This model consists of the following systems and libraries:
\begin{itemize}
	\item Library \texttt{A} has 1 version \lib{A}{1}.
	\item System \texttt{B} has 2 versions \sys{B}{1} and \sys{B}{2}.
	\item Library \texttt{C} has 2 versions \lib{C}{1} and \lib{C}{2}.
	\item System \texttt{D} has 3 versions \sys{D}{1}, \sys{D}{2} and \sys{D}{3}.
\end{itemize}
Figure \ref{fig:migration} depicts the following library dependency relationships as an orange dotted line. Below we list all dependencies between these systems and libraries at some point in time:
\begin{itemize}
	\item Library \lib{A}{1} is used as a dependency of system \texttt{B}.
	\item Library \lib{C}{1} is used as a dependency of system \texttt{B} and \texttt{D}.
	\item Library \lib{C}{2} is used as a dependency of system \texttt{D}.
\end{itemize}

From a system perspective, our model is able to track how often maintainers update their libraries.
Since a system version may contain multiple dependency migrations, we track the number of migrations that occur during one system update, which is denoted as \dmc.

\begin{quote}
	\textit{\dmcfull} (\textbf{\dmc}) is a count of library migrations that occur at one system version update.
\end{quote}

Figure \ref{fig:migration} depicts an example of a \dmc~update where at the release of \sys{B}{2}, one dependency update occurred (\ie~\dmc=1).
We can see in the figure, that for \sys{B}{2}, a new dependency (\lib{A}{1}) is added while still keeping the \lib{B}{1} dependency. 

From the alternative library viewpoint, our model is able to track library usage trends over time.
We track the number of library migrations that occur within the universe of known systems to determine the usage of a library, which is denoted as LU.
\label{sec:LU}
\begin{quote}
\textit{Library Usage} (\textbf{LU}) is the total population count of dependent systems at a specific point in time. 
\end{quote}

Figure~\ref{fig:migration} shows an example of the LU metrics. The figure shows that at $x$ point in time, the LU of \lib{C}{1} is two (\texttt{B} and \texttt{D}).
However at point $y$, since \sys{D}{2} migrates its dependency to \lib{C}{2}, the LU of \lib{C}{1} becomes one (\texttt{B}) while the LU of \lib{C}{2} is now one (\texttt{D}).
Moreover, systems can depend on older versions of a library.
This is modeled and shown in the figure, as a line branching out from the original line of libraries.
For instance, library \texttt{C} separates into two different branches because \lib{C}{1} is still being actively depended upon by other systems (\ie~\sys{A}{2}).


 \begin{table}[t]
 	\begin{center}
 		\tabcolsep=0.1cm
 		\caption{ Summary of System and Library migration metrics defined for (RQ1). Note we use dep. = Dependencies and ver. = version}
 		\label{tab:RQ1metrics}
 		\begin{tabular}{l|c|l|rc}
 			\hline
 			\multirow{1}{*}{\textbf{Alias}}
 			&\multirow{1}{*}{\textbf{Dimension}}
 			& \multirow{1}{*}{\textbf{Metric}}
 			& \multirow{1}{*}{\textbf{Brief Description}}
 			\\
 			\textbf{m1}&System &Dep. Per System (\#Dep.)& \# Dependencies \\
 			\textbf{m2}&&Dep. Update Per System (\dmc)& \# Dependencies updated \\\hline
 			\textbf{m3}&Library &Library Usage(LU)& \# library users \\
 			\textbf{m4}&&Peak LU & max. \# library users \\
 			\textbf{m5}&&Current LU & current \# library users\\
 			\textbf{m6}&&Pre-Peak& time to reach Peak LU \\
 			\textbf{m7}&&Post-Peak&time after Peak LU \\
 			\textbf{m8}&&Library Residue& \% remaining systems after Peak LU\\\hline
 		\end{tabular}
 	\end{center}
 \end{table}

 \begin{figure}
 	\centering
 	\includegraphics[width=0.6\textwidth]{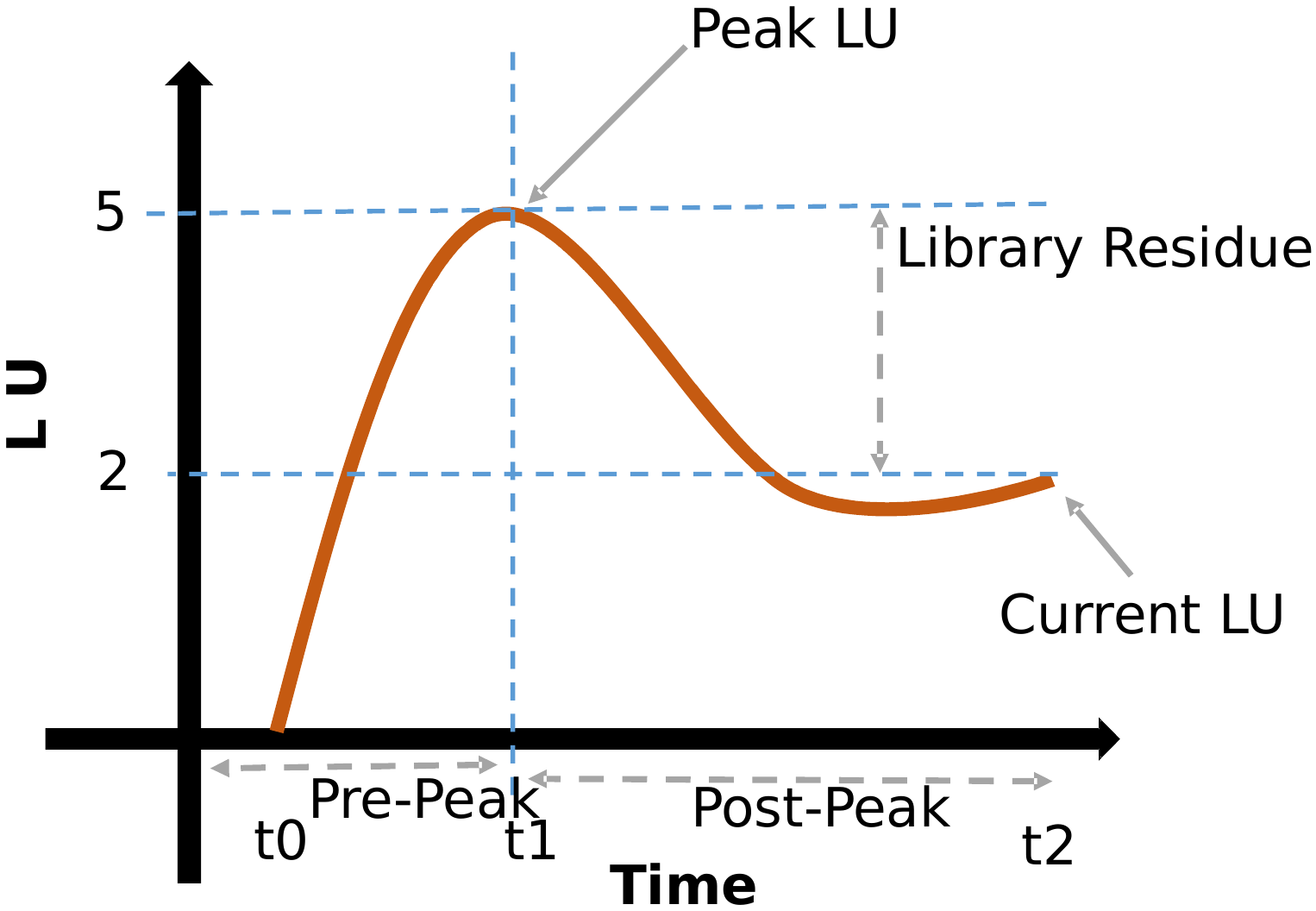}
 	\caption{Simple example of the LU-based metrics. We show the Peak LU at time t1, current LU at time t2 and library residue (Peak LU / Current LU).}
 	\label{fig:LUmetrics}
 \end{figure}

\paragraph{\textbf{(2) Analysis Method}:}
Table \ref{tab:RQ1metrics} provides a summary of the metrics provided by our model. 
To fully understand this phenomena, we analyze library migrations from both the system and library dimensions.

From the system dimension, we use system metrics to investigate the distribution of dependencies per system (\texttt{m1}) and the frequency of library migrations per library (\texttt{m2}).
First, we utilize boxplots and descriptive statistics to report the median ($\bar{x}$) and mean ($\mu$) for each metric.
We then test the hypothesis that \textit{systems with more dependencies tend to have more frequent updates}.
We employ the Spearman and Pearson correlation tests \citep{Edgell84} to determine any correlation relation between metrics \texttt{m1} and \texttt{m2}.
A high correlation score confirms the assumption that a more complex systems will tend to have more updates, while a low correlation will confirm the hypothesis that \textit{the number of library dependencies does not influence the frequency of updates}.

\sloppypar{
From the library dimension, we investigate how the migration away from a specific library dependency spreads over time. 
This work is inspired by the Diffusion of Innovation curves \citep{DoI}, which seeks to explain how, why, and at what rate new ideas and technology spreads.
Figure \ref{fig:LUmetrics} is a visual example of the LU metrics from Table \ref{tab:RQ1metrics}.
We utilize the LU metrics to study the (i) LU trends (\ie~whether or not a library dependency is gaining or losing system users) and the (ii) rate of decline after system users begin to migrate away from the dependency.
Based on the LU (\texttt{m3}) metric, Figure \ref{fig:LUmetrics}  introduces a simple example of the derived LU metrics that characterize a LU trend:
}

\begin{itemize}
	\item \textit{LU counts} - The \textit{Peak LU} (\texttt{m4}) metric describes the maximum population count of user systems reached by a dependency.
	The \textit{Current LU} (\texttt{m5}) is a related metric that describes the latest population count of user systems that actively use this dependency in their systems.
	
	\item \textit{LU over time} -  The \textit{Pre-Peak} (\texttt{m6}) metric refers to the time taken for a dependency to reach a peak LU (days).
	Conversely, \textit{Post-Peak } (\texttt{m7}) metric  refers to the time passed since the peak LU was reached (days).
	
	\item \textit{LU rate after Peak LU} - The \textit{Library Residue} (\texttt{m8}) metric describes the percentage of user systems remaining after \textit{Peak LU} (\texttt{m4}) has been reached for a dependency (\ie~Current LU (\texttt{m5}) / Peak LU (\texttt{m4})).
\end{itemize}

In Figure \ref{fig:LUmetrics}, we show the LU metrics as a LU trending curve.
In detail, we find that the \textit{Peak LU} is 5 users at t1, with the \textit{current LU} at 2 users.
At the starting point $t0$, \textit{Pre-Peak} is the period from $t0$ to $t1$ and \textit{Post-Peak} being the time from $t1$ to $t2$.
Quantitatively, we conjecture that the low Library Residue (\ie~40\% (2/5)) indicates that a developer using this dependency should consider migration towards a replacement dependency.

To address the library dimension of (RQ1), we present four statistical analysis to report the LU trends.
First, we use a cumulative frequency distribution graph to understand the distribution of popular library versions (\texttt{m4} and \texttt{m5}).
We then use a cumulative distribution to measure the average time for libraries to reach their peak usages (\texttt{m6} and \texttt{m7}).
Third, we use boxplots to measure the distribution of the Library Residue metric (\texttt{m8}).
Finally, we plot and analyze the amount of system dependencies and their Library Residue. 

\paragraph{\textbf{(3) Data Collection}:}
It is important that we test our approach from a quality set of real-world projects to improve confidence on our results.
Therefore, we conducted a large-scale empirical evaluation of software systems and library migrations, focusing on popular Java projects that use Maven libraries as their third-party dependencies.
We mine and collect projects that reside in GitHub\footnote{\url{https://github.com/}} as the source of our dataset.
To ensure that our dataset is a quality representation of real-world applications, we enforce the following pre-processing data quality filters:

\begin{itemize}
	\item \textit{Projects that are mature and well-maintained - }
	 The first quality filter is to ensure that migrations are indicative of active and large-scale projects that are hosted on GitHub (\ie~removing toy projects).
	 Hence, we select projects that had more than 100 commits and had at least a recent commit between January 2015 and November 2015.

	\item \textit{Projects that are unique and not duplicates - }
	The second quality filter is to ensure that no duplicates exist within the collected dataset.
	Hence, we semi-automatically inspect repository names to validate that none of the projects are forks from other projects (\ie~same project name in different repository).

	\item \textit{Projects that use a dependency management tool - }
	We conjecture that projects managed by a dependency management tool is more likely to consider library migration practices.
	Therefore, the third filter distinguishes projects that implement a dependency management tool such as the Maven dependency management tool.
	For a Maven dependency, every project in the Maven repository includes a Project Object Model file (\ie{} \texttt{pom.xml}) that describes the project's configuration meta-data ---including its compile and run time dependencies.

		\begin{lstlisting}[language=XML,
		caption={Code snippet of the \texttt{pom.xml} metafile for the \texttt{GitWalker} system showing the dependency relationship to between two Maven dependencies, \texttt{javaparser} and \texttt{gitective-core}.}, label=list:pom, frame=single]
		...
		<groupId>GitWalker</groupId>
		<artifactId>GitWalker</artifactId>
		<version>0.0.1-SNAPSHOT</version>
		<name>GitWalker</name>
		...
		<dependencies>
			<dependency>
				<groupId>com.google.code.javaparser</groupId>
				<artifactId>javaparser</artifactId>
				<version>1.0.8</version>
			</dependency>
			<dependency>
				<groupId>org.gitective</groupId>
				<artifactId>gitective-core</artifactId>
				<version>0.9.9</version>
			</dependency>
		</dependencies>
		...\end{lstlisting}

	Listing \ref{list:pom} shows a \texttt{pom.xml}, which lists dependency relationships between a particular system version with any valid Maven library version.
	In this example, we extract the dependency relation for system \sys{Gitwalker}{0.0.1-SNAPSHOT} that uses the \lib{javaparser}{1.0.8} and \lib{gitective-core}{0.9.9} dependencies.
	To automatically extract the history of dependency migrations for a project, we mine the historic changes of the \texttt{pom.xml}.
	We package our method in a tool called PomWalker\footnote{\url{https://github.com/raux/PomWalker}}.

	\item \textit{Popular and latest dependency versions - } LU trends require sufficient usage by systems.
	As a result, we focus on the more popular libraries for a higher quality result. 
	Moreover, to capture migrations away from a library dependency, we filter out the latest versions of any library in the dataset.
\end{itemize}

 \begin{table}[t]
 	\begin{center}
 		\fontsize{8}{10}\selectfont
 		\tabcolsep=0.1cm
 		\caption{ Summary of the collected dataset}
 		\label{tab:dataset}
 		\begin{tabular}{lcccc}
 			\hline
 			\multirow{1}{*}{}
 			& \multirow{1}{*}{\textbf{Dataset statistics}}
 			\\
 			projects creation dates &2004-Oct to 2009-Jan& \\
 			projects last update & 2015-Jan to 2015-Nov\\  \hline
 			\# unique systems (projects) &48,495 (4,659)& \\
 			\# unique library versions & 2,736& \\
 			total size of projects & 630 GB & \\
 			\hline
 			\# commits related to \texttt{pom.xml} &4,892,770& \\
 			\# library dependency migrations & 852,322 \\  \hline
 		\end{tabular}
 	\end{center}
 \end{table}

Table \ref{tab:dataset} presents a summary of the filtered 4,659 projects after pre-processing from an original collection of 10,523  GitHub projects.
Our study tracks dependency migration between a Maven library and each unique system within each project (\ie~a project may contain multiple systems). 
We then mine 48,495 systems from the 4,659 software projects to extract 852,322 dependency migrations.
For the LU trend analysis, we filter out rarely used libraries (\ie~dependencies with less than 4 user systems are defined as unpopular) and 213 of the latest library versions, leaving 2,736 library versions available for our study.

\subsection {(RQ2) \RqTwo}
Our method to answer the second research question (RQ2) is through a case study analysis of developer responsiveness to the awareness mechanism.
It is comprised of three steps: (1) tracking library migration in response to awareness mechanisms (2) analysis method (3) data collection.
Case studies for the new release announcement are presented in Section \ref{sec:nr}, with those for the security advisory presented in Section \ref{sec:ad}.

 \begin{figure}
 	\centering
 	\includegraphics[width=1\textwidth]{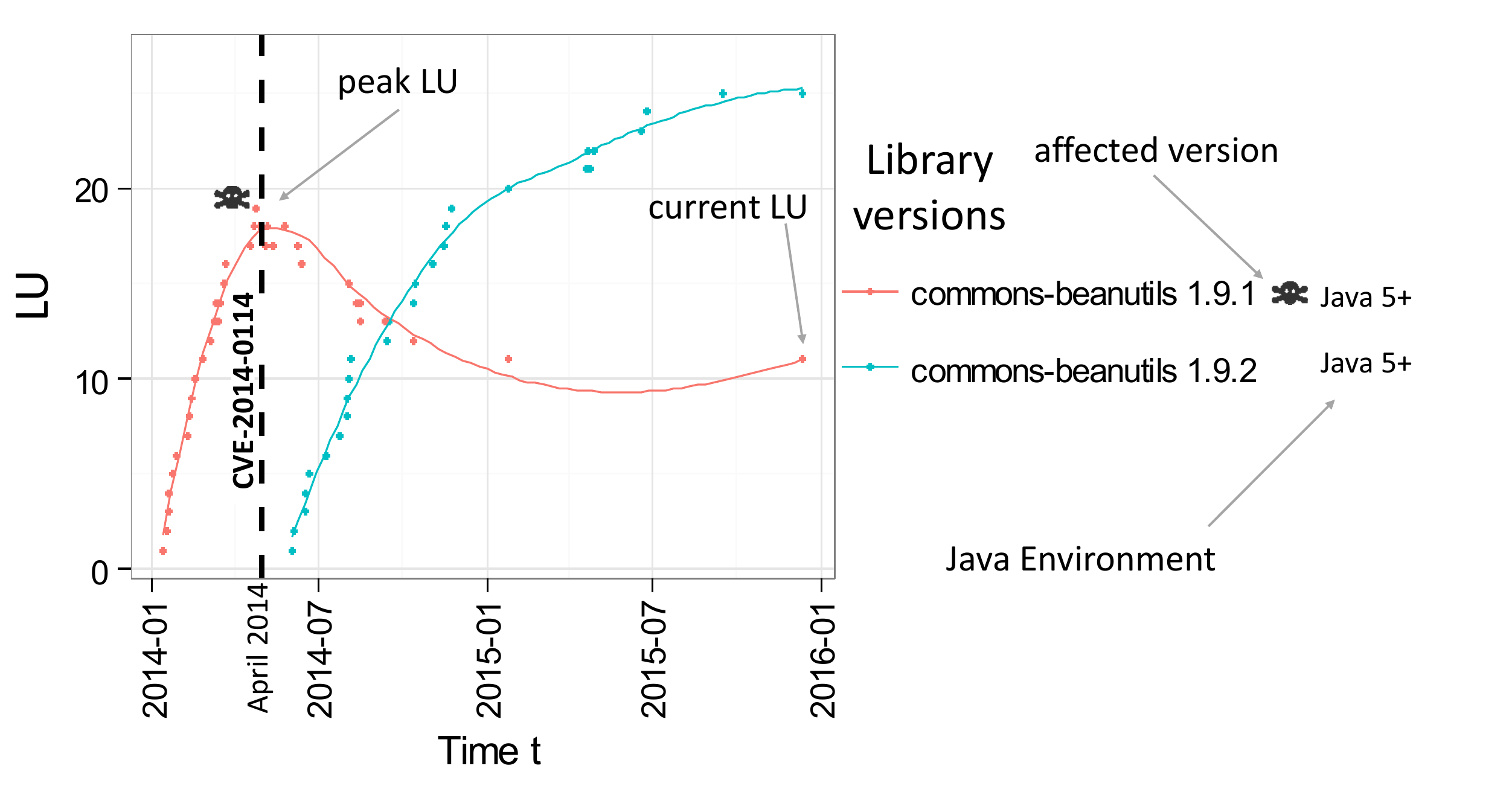}
 	\caption{A Library Migration Plot for libraries \lib{beanutils}{1.9.1} and \lib{beanutils}{1.9.2}.
 		In this example, the release of a related security advisory \texttt{CVE-2014-0114} (black dashed line) that affects \lib{beanutils}{1.9.1} (marked with crossbones). We also show which JDK (5+) version in which the version supports.}
 	\label{fig:LAP}
 \end{figure}

\paragraph{\textbf{(1) Tracking Migration in Response to Awareness Mechanisms}:}
Figure \ref{fig:LAP} presents the Library Migration Plot (LMP) used to track LU trends over time.
Together with documentation, we use LMPs to infer library migration patterns and trends.
The LMP shows LU changes in the library (y-axis) with respect to time (x-axis).
The LMP curve itself should not be taken at face value, as the smoothing algorithm is generated by a predictive model and it is not a true reflection of all data points.
In Figure \ref{fig:LAP}, we observe that the \texttt{commons-beanutils} library \lib{commons-beanutils}{1.9.1} (red line) had 19 user systems using it as a dependency in April 2014.
Then by January 2015, its LU had decreased to 11 user systems.
The LMP depicts an effect of awareness mechanisms through annotation of either  or a new release announcement or a security advisory as follows:
\begin{itemize}
\item \sloppypar{\textit{Official Release Announcement} - Figure \ref{fig:LAP}  depicts an example of two versions: \lib{commons-beanutils}{1.9.1} and  \lib{commons-beanutils}{1.9.2}. Hence, we can use the LMP to compare the migration patterns between versions of a library. For instance, the LMP presents the effect of the new release of \lib{commons-beanutils}{1.9.2}, illustrated by the declining LU curve at \lib{commons-beanutils}{1.9.1}}.

\item  \sloppypar{\textit{Security Advisory Disclosure} - Figure \ref{fig:LAP} annotates when the security advisory \texttt{CVE-2014-0114} was disclosed to the public (\ie~April 2014). In detail, the LMP presents evidence of how a security vulnerability triggers the library migration from \lib{commons-beanutils}{1.9.1}, illustrated by its declining LU curve.}
\end{itemize}

\paragraph{\textbf{(2) Analysis Method}:}
Our approach to answer (RQ2) involves a manual case study analysis to understand developer responsiveness to either a new release announcement or a security advisory.
For more useful and practical scenarios, selection of our case studies included (i) new releases from the more popular libraries (\ie~as they tend to impact more developers) and (ii) more severe security advisories  (\ie~warrants immediate developer attention).

At the quantitative level, we first visually analyze the LMP, using our LU metrics to quantify the LU trend response towards the awareness mechanism.
We then manually consult online documentation such as the release logs, and its semantic versioning schema to estimate the effort needed to migrate towards a newer replacement dependency. 
For the vulnerable dependencies, we consult information from the security advisory and the life-cycle of a vulnerability (See Section \ref{sec:am}) to estimate the needed migration effort.
For example, in Figure \ref{fig:LAP}, we infer from the release notes that \lib{commons-beanutils}{1.9.1} to \lib{commons-beanutils}{1.9.2} update is a compatible minor update with 2 bug fixes and 1 new feature.
Since both are supported by the latest JDK (Java 5 and higher), we assume that the migration effort required is much lower compared to a migration to a different JDK environment.

\begin{table}[t]
	\begin{center}
		\fontsize{8}{10}\selectfont
		\tabcolsep=0.1cm
		\caption{Top 20 LU library versions
		}
		\label{tab:newCandid}
		\begin{tabular}{llccc}
			\hline
			\multirow{1}{*}{}
			& \multirow{1}{*}{\textbf{Library}}
			& \multirow{1}{*}{\textbf{Versions}}
			\\
			*&junit& (4.11), (4.10), (4.8.2), (3.8.1), (4.8.1)\\
			&javax.servlet-servlet-api& (2.5)\\
			&commons-io-commons-io& (2.4), (2.6)\\
			*&log4j-log4j&(1.2.16), (1.2.17)\\
			&commons-lang& (2.6)\\
			&commons-logging& (1.1.1)\\
			&commons-lang& (3-3.1)\\
			&commons-collections& (3.2.1)\\
			&javax.servlet-jstl& (1.2)\\
			&org.mockito-mockito-all& (1.9.5)\\
			&commons-httpclient&(3.1)\\
			*&guava&(14.0.1), (18.0)\\
			&commons-dbcp& (1.4) \\ \hline
		\end{tabular}
	\end{center}
\end{table}

\begin{table}[t]
	\begin{center}
		\fontsize{8}{10}\selectfont
		\tabcolsep=0.1cm
		\caption{New Release case studies from three popular libraries. For each library, we look at the LU trends of three libraries.
		}
		\label{tab:newVer}
		\begin{tabular}{lcccc}
			\hline
			\multirow{1}{*}{\textbf{Alias}}
			& \multirow{1}{*}{\textbf{Library}}
			& \multirow{1}{*}{\textbf{ver.1}}
			& \multirow{1}{*}{\textbf{ver.2}}
			& \multirow{1}{*}{\textbf{ver.3}}
			\\
			NR1&{google-guava}  &16.0.1 (2014-02-03) & 17.0 (2014-04-22) &18.0 (2014-08-25)\\
			NR2&{junit}  &3.8.1 (2002-08-24) & 4.10 (2011-09-29) & 4.11 (2012-11-15)\\
			NR3&{log4j}  &1.2.15 (2007-08-24) & 1.2.16 (2010-04-06) & 1.2.17 (2012-05-06) \\ \hline
		\end{tabular}
	\end{center}
\end{table}

\paragraph{\textbf{(3) Data Collection}:}
Since our research method to answer (RQ2) is through the use of case studies, we systematically select a subset of eligible projects from the dataset collected in (RQ1).
Selection of a new release candidate is comprised of three steps.
First, since our objective is to find common LU trends popular libraries, we select the top 20 library versions out of the 2,736 libraries.
The top 20 libraries are shown in Table \ref{tab:newCandid}.
Then, for each of the 20 library versions, we generate and categorize them based on LMP curve patterns.
Finally, we select three case studies that depict distinctive LU trends.
Table \ref{tab:newVer} shows the nine popular library versions of  \texttt{google-guava}\footnote{\url{https://code.google.com/p/guava-libraries/}}, \texttt{junit}\footnote{{\url{http://junit.org/}}} and \texttt{log4j}\footnote{{\url{http://logging.apache.org/log4j/1.2/}}} that meet our selection criteria.

\begin{table*}[t]
	\begin{center}
		\fontsize{7}{12}\selectfont
		\tabcolsep=0.05cm
		\caption{Security Advisory case studies from the Apache Family of Maven libraries. Note that the affected versions include all prior versions. Likewise safe versions also include all superseding versions.
		}
		\label{tab:vulnerableDataset}
		\begin{tabular}{lccccc}
			\hline
			\multirow{1}{*}{\textbf{Alias}}
			& \multirow{1}{*}{\textbf{CVE Id}}
			& \multirow{1}{*}{\textbf{library}}
			& \multirow{1}{*}{\textbf{Release}}
			& \multirow{1}{*}{\textbf{Affected ver.}}
			& \multirow{1}{*}{\textbf{Attack(CVSS)}}
			\\
			V1&CVE-2014-0114 & commons-beautils & 2014-04-30 & 1.9.1 & Denial of Service (7.5)\\
			V2&CVE-2014-0050 & commons-fileupload & 2014-01-04& 1.3 & man--in--the--middle(5.8)\\
			V3&CVE-2012-5783 &  commons-httpclient & 2012-04-11 &3.x & man--in--the--middle(4.3)\\
			V4&CVE-2012-6153 & httpcomponents & 2014-09-04& 4.2.2 & man--in--the--middle(7.5)\\
			V5&CVE-2012-2098 & commons-compress & 2012-06-29 &1.4 & man--in--the--middle(5.0)\\ \hline
		\end{tabular}
	\end{center}
\end{table*}

Table \ref{tab:vulnerableDataset} shows the 5 security advisory case studies that meet our selection criteria.
As part of the selection criteria process, we manually inspect and match CVE security advisories between 2009-2014, that affected any of our collected systems in (RQ1).
Particularly, we select 123 products from the popular Apache Software Foundation (ASF) products, and associated with 686 disclosed security advisories\footnote{An updated listing is available online at \url{http://www.cvedetails.com/product-list/vendor_id-45/apache.html}}.
We find that 15 out of the 123 ASF products were third-party libraries. 
We then select case studies that had severe risk of malicious exposure to attackers and would require immediate attention of the developer.  
Specifically, the security advisory should have a medium to high Common Vulnerability Score (CVSS)\footnote{it is officially known as the CVSS v2 base score. The calculation is shown at \url{https://www.first.org/cvss/v2/guide}} (\ie~4 or higher). 
So out of the remaining 15 libraries, we select 5 security advisory cases with a CVSS base score of 4 or higher.
As shown in Table \ref{tab:vulnerableDataset} our selected case studies exhibit the following malicious exposures: \texttt{V1} causes a \textit{Denial of Service (DoS)} with a high CVSS score.
The remaining four security advisory cases all describe web application exposure to a remote \textit{`man in the middle'} web attack, with a medium-to-high CVSS severity rating.

\subsection {(RQ3) \RqThree}
Our research method to answer the third research question (RQ3) is through a survey targeting developers that belong to projects that were identified as non responsive to a severe security advisory.
The method comprises of two steps: (1) survey design and (2) data collection.
Results to (RQ3) are presented in Section \ref{sec:barriers}.

\paragraph{\textbf{(1) Survey Design}:}
Our research method makes use of a qualitative survey interview form.
Listing \ref{list:email} shows the template of our survey form\footnote{the complete form is available at \url{http://sel.ist.osaka-u.ac.jp/people/raula-k/librarymigrations/questionaire.html}} sent to developers of the contactable projects. 
Not all projects facilitate a contact medium, so we targeted projects that allowed public communication, either through an issue management system or a mailing list.
The survey form is designed with two parts.
First, we customize the survey form to include project specifics, such as the exact location of the \texttt{pom.xml} file where the dependency is being relied upon by the project.
We then ask developers to respond on the following two questions: (i)\textit{Were you aware of the vulnerability? If so, then how long ago} and (ii) \textit{What are some factors that influence you not to update?} 

\begin{minipage}{\linewidth}
\begin{lstlisting}[language= email,
basicstyle=\ttfamily\small,
columns=fullflexible,
showstringspaces=false,
commentstyle=\color{gray}\upshape,
numbers=none,
numbersep=5pt,
showspaces=false,
showstringspaces=false,
showtabs=false,
caption={Email snippet of the survey form sent to developers of the selected projects that were non responsive to a security vulnerability.}, label=list:email,
tabsize=2, frame=lrtb]
<!---email snippet/>
Dear GitHub OSS Developer,
...
As a part of my study I particularity focused on the <library version/> and
the <CVE-xxx-xxxx/> <CVE URL/>, announced on <date>, which affects versions xxx.
We noticed that your project on GitHub is still configured to depend on a
vulnerable version of <library version/> at  <https://xxx.xxx.xx/pom.xml/>
We understand that there are many reasons for not migrating, thus we appreciate
if you could simply detail the following:
1. Were you aware of the vulnerability? If so, then how long ago.
2. What are some factors that influence you not to update?
...
<!---email snippet/>\end{lstlisting}
\end{minipage}

For the analysis, we first tally responses according as to whether or not the developer was aware of the vulnerable threat.
Our strategy to analyze the feedback is through a systematic (i) reading of each response, (ii) checking and summarizing text by consistency and omissions and (iii) looking for similarities or differences between interviewee responses.
We perform the analysis in three steps. 
First, the main author performs a categorization of responses.
Then, another author is tasked to verify and criticize each category of responses.
Finally, the categories are presented to rest of the authors for a group consensus.   

\paragraph{\textbf{(2) Data Collection}:}
Since our approach to answer (RQ3) is through a survey, our data is from the security advisory case studies in (RQ2).
From the LMP analysis in (RQ2), we identified candidate projects that are non responsive to the security advisory announcement.
Since we collected 16 developer responses, categorization of the similarities and differences was manageable by one author and then later criticized and verified by other authors for the final consensus.
All results of the collected dataset, including the tally of listed and contactable projects are presented in Section \ref{sec:barriers}.

\section{Library Migration in Practice}
\label{sec:prac}

In this section, we present the results for (RQ1) \RqOne
In detail, we present the statistical results from both a system (Section \ref{sec:sys}) and library dimension (Section \ref{sec:lib}), before finally answering (RQ1).

\begin{figure*}[t]
	\centering
	\subfigure[ \# dependencies per system ]{\label{fig:sysLU}%
		\includegraphics[width=0.3\columnwidth]{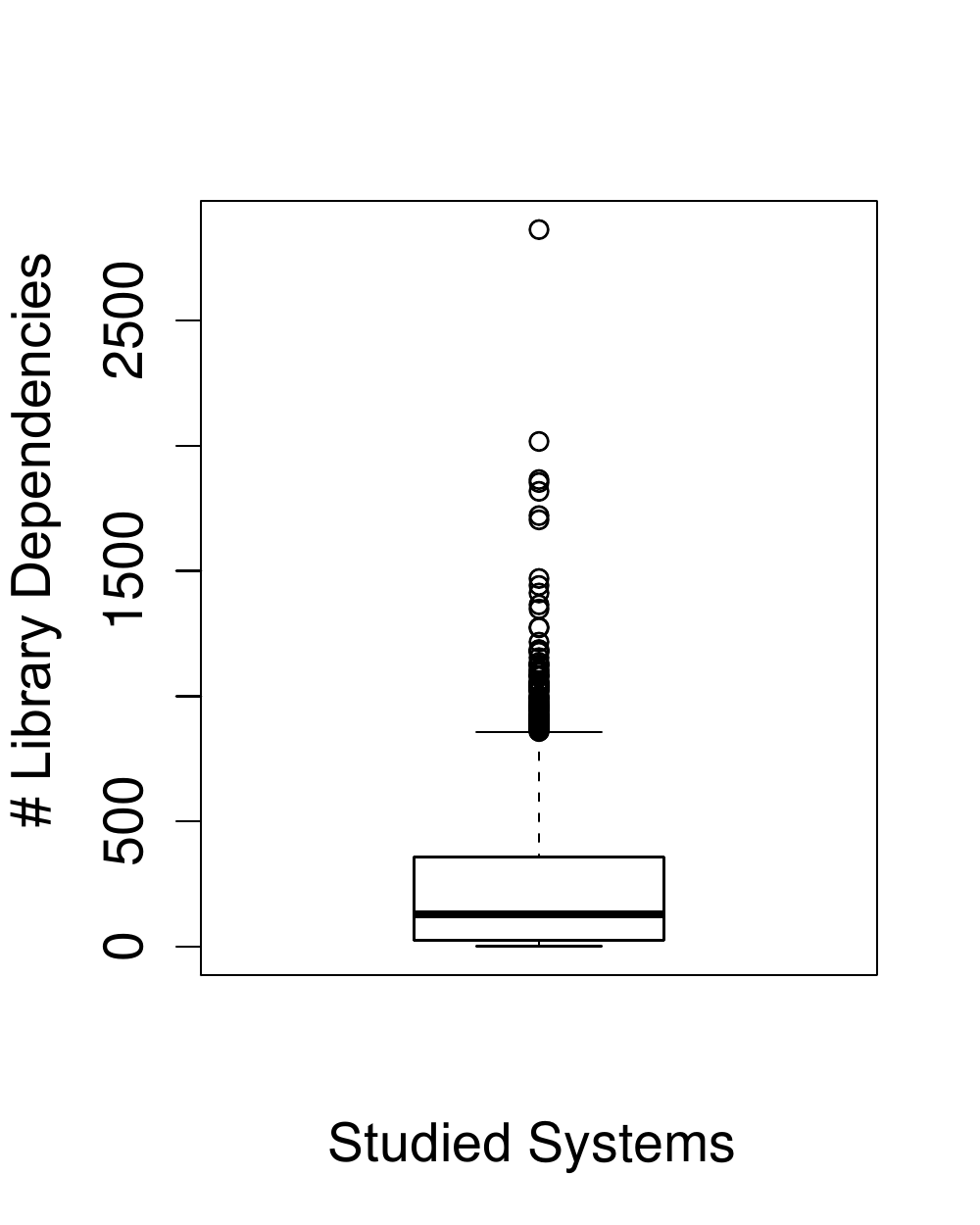}
	}
	\subfigure[\# \dmc~per system]{\label{fig:fLU}
		\includegraphics[width=0.3\columnwidth]{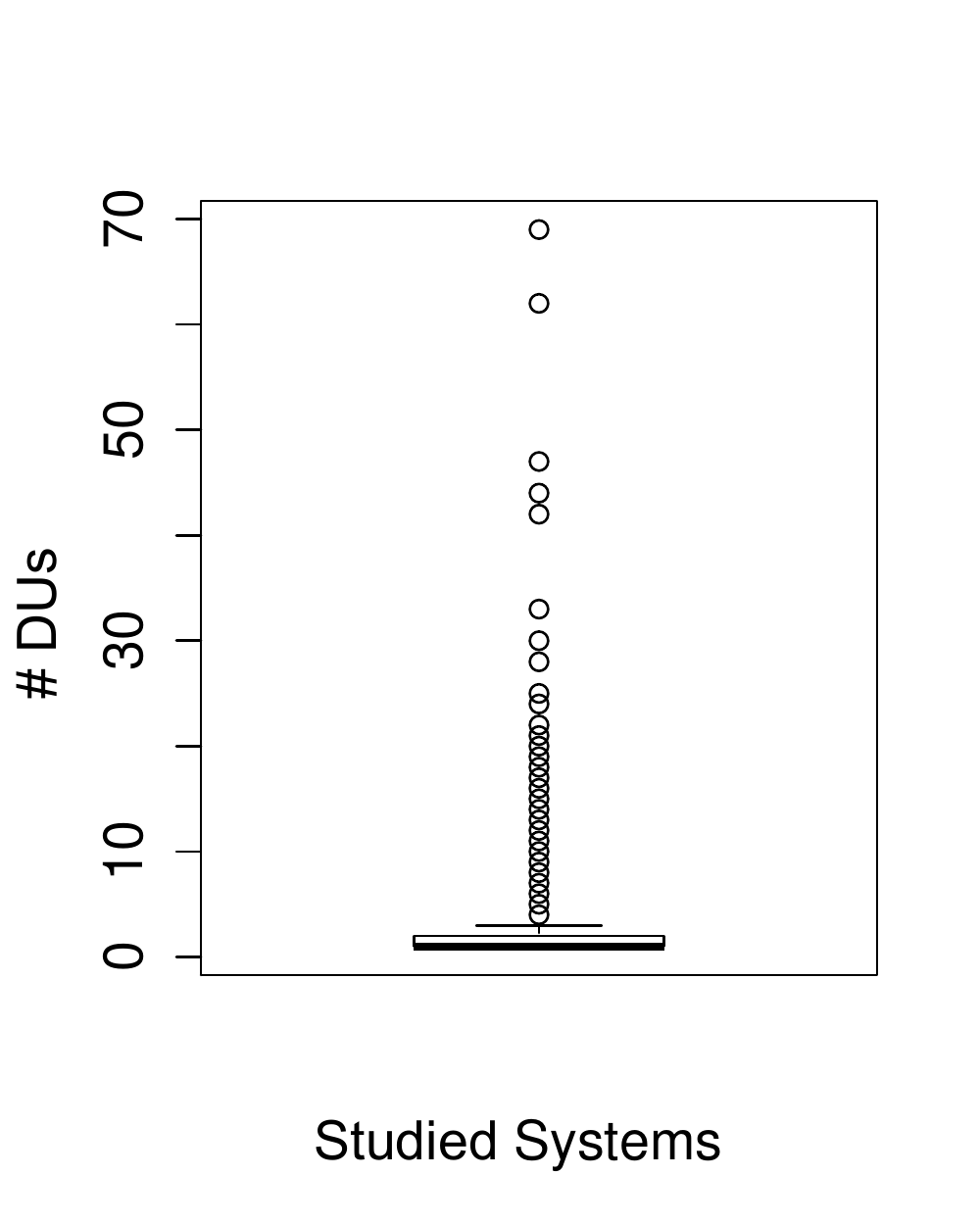}
	}
	\subfigure[\# dependencies vs. \# \dmc]{\label{fig:freqvsDeps}
		\includegraphics[width=0.3\columnwidth]{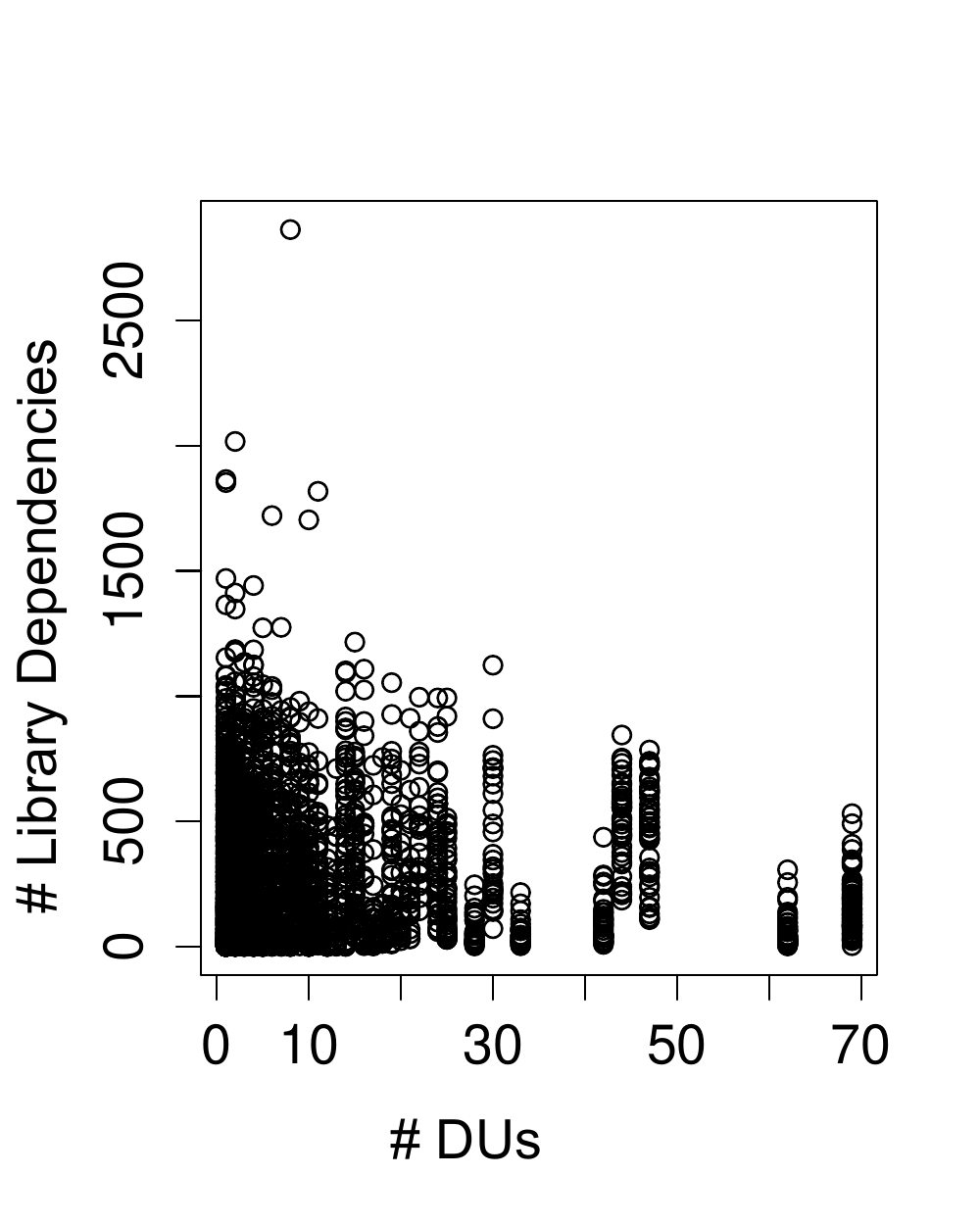}
	}
	\caption{Updates from a System dimension depicts (a) \# of dependencies per system. ($\bar{x}$=147, $\mu$=267.2, $\sigma$=311.56) (b) frequency of \dmc s per system ($\bar{x}$=1, $\mu$=2.4, $\sigma$=4.2) and (c) relationship between \# of dependencies vs. \# of \dmc s (log-scale). }
	\label{fig:sysStats}
\end{figure*}

\subsection{System Dimension}
\label{sec:sys}

Figure \ref{fig:sysStats} shows the results on how maintainers manage and update their dependencies from a system viewpoint.
Specifically, the distribution of library dependencies per system in Figure \ref{fig:sysLU} confirms that systems show heavy dependence on libraries ($\bar{x}$=147, $\mu$=267.2, $\sigma$=311.56).
A reason for this heavy reliance on libraries is because many of the analyzed projects are comprised of multiple subsystems which form a complex set of dependencies.
Furthermore, Figure \ref{fig:fLU} suggests that systems rarely update library dependencies, with a low frequency of \dmcfull s per system (i.e., $\bar{x}$=1, $\mu$=2.4), with each \dmc~containing at least two library dependencies (i.e., $\bar{x}$=2, $\mu$=4.1, $\sigma$=14.9).
Finally, according to Figure \ref{fig:freqvsDeps} visually, we did not find a strong correlation between the number of library dependencies and the frequency of \dmc, with statistical tests reporting weak correlations (pearson = 0.05, spearman = 0.07). This result confirms the hypothesis that the \textit{number of library dependencies in a system does not influence the frequency of updates}.

\begin{figure*}[t]
	\centering
	\subfigure[LU Distributions per dependency]{\label{fig:PLU}%
		\includegraphics[width=0.5\columnwidth]{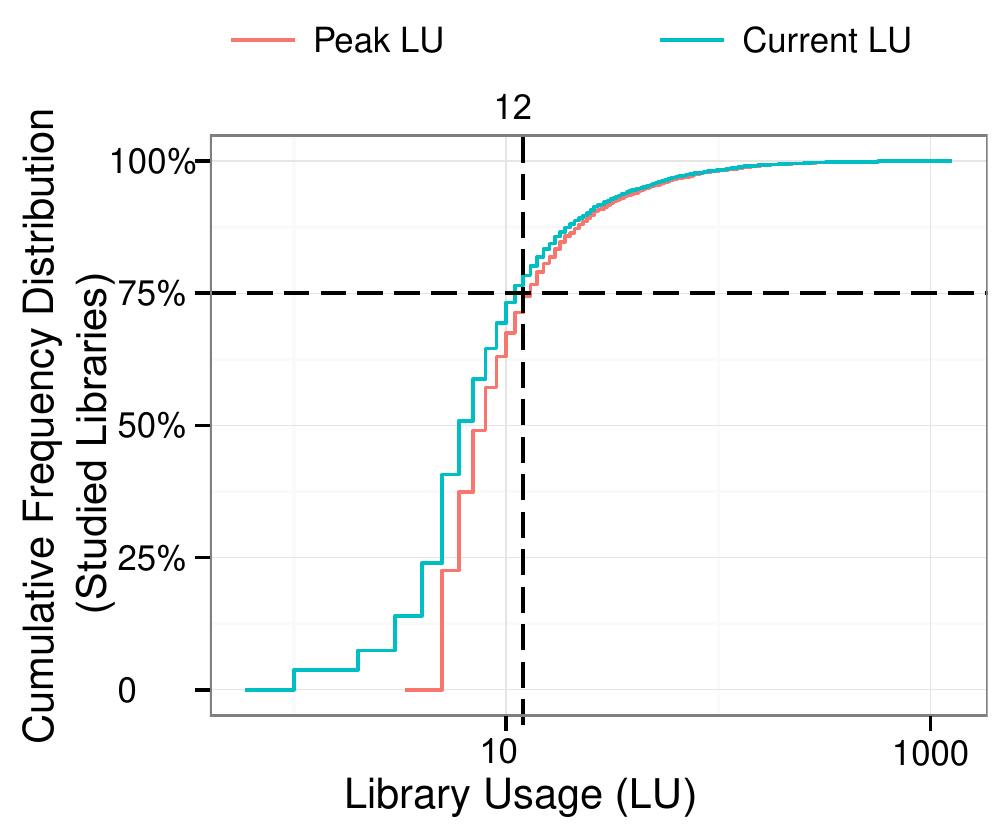}
	}
	\subfigure[Time-frame analysis]{\label{fig:TP}
		\includegraphics[width=0.5\columnwidth]{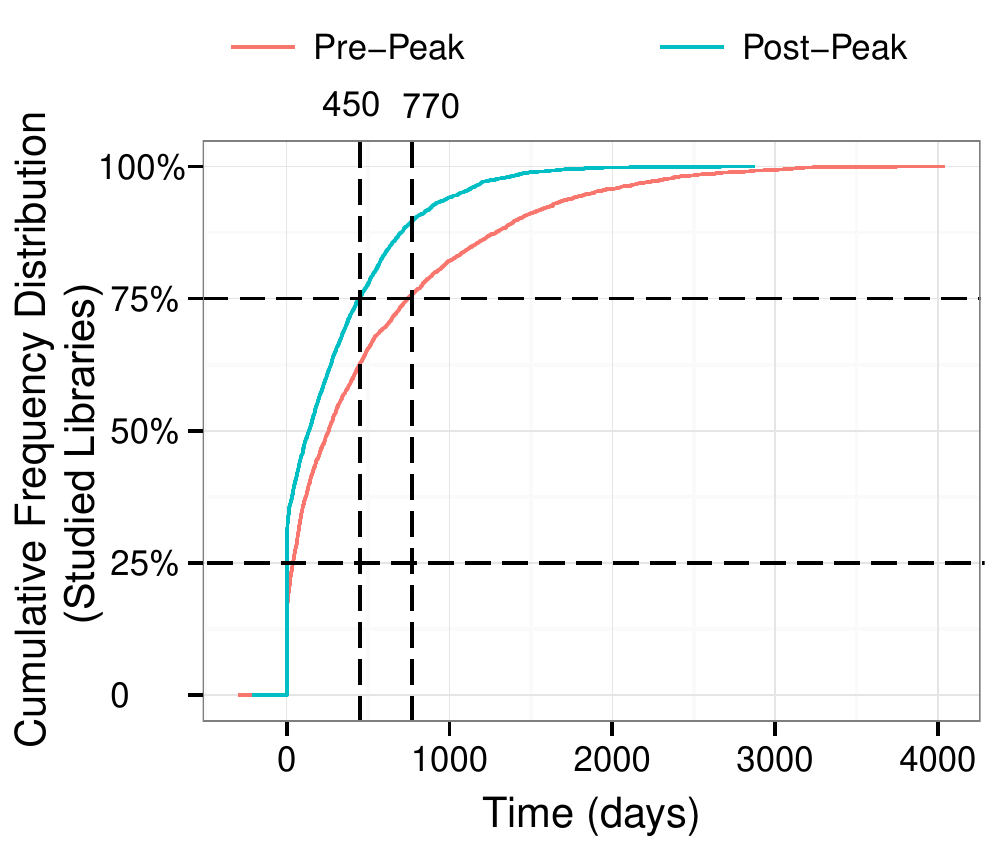}
	}
	\subfigure[Library Residue per dependency]{\label{fig:LR}
		\includegraphics[width=0.35\columnwidth]{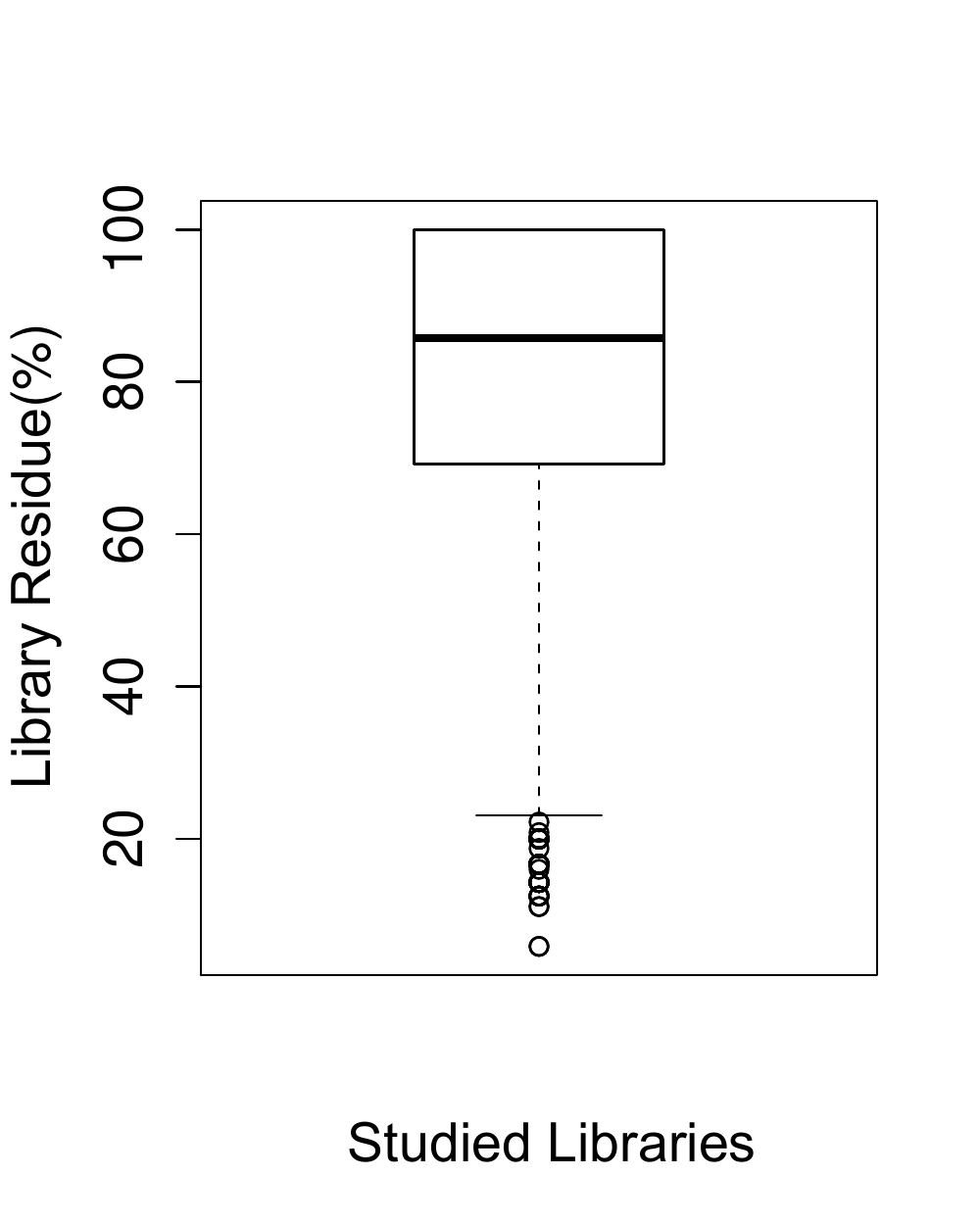}
	}
	\caption{Updates from a Library dimension depicts the cumulative frequency distribution (a) of Peak LU and Current LU  (Log scale), (b) time-frame metric distributions and the boxplot of (c) library residue (\%) for 2,736 dependencies.}
	\label{fig:RQ1}
\end{figure*}


\begin{figure}[t]
	\centering
	\includegraphics[width=.5\textwidth]{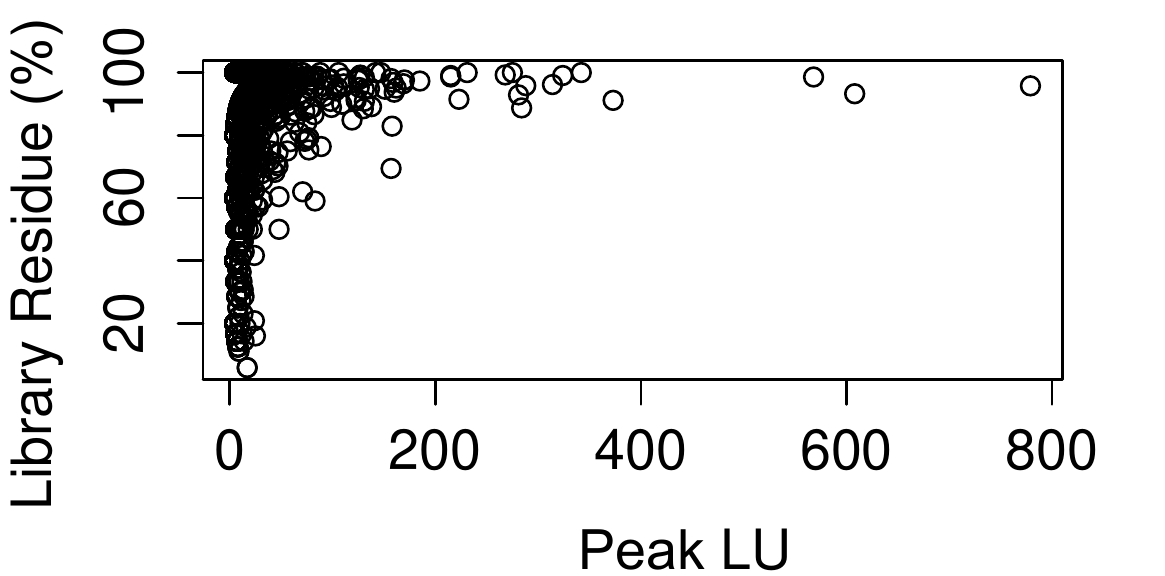}
	\caption{A correlation of Library Residue against Peak LU, showing that popular library dependencies (with higher peaks) also tend to exhibit higher Library Residue. }
	\label{fig:residueVsPLU}
\end{figure}

\subsection{Library Perspective}
\label{sec:lib}
Figure \ref{fig:RQ1} and Figure \ref{fig:residueVsPLU} presents LU trends of library dependencies used by our studied systems.
Figure \ref{fig:PLU} shows that LU for 75\% of the popular libraries is 12 (\ie~peak LU).
Interestingly, we also found 596 libraries that exhibited no library migration, such that peak library usage is the current library usage (i.e., peak LU=current LU).
Additionally, Figure \ref{fig:TP} shows that reaching the peak library usage is slow for most dependencies.
Concretely, the figure shows that 25\% of dependencies took less than a day to reach their peak LU.
Afterwards the rate slows down (depicted by curve), showing 75\% of dependencies took less than 770 days to reach their peak LU (i.e., Pre-Peak).
Upon closer inspection, we found that these dependencies were specialized libraries that were used by a smaller number of systems (\ie~low LU).
After reaching peak usage, dependent systems tend to slowly migrate away, shown in Figure, with 75\% of dependencies experiencing some migration of its users over the next 450 days (ie., Post-Peak).
Importantly, Figure \ref{fig:LR} suggests that many systems remain with an outdated dependency, even after some library migration away from the dependency has begun.
The figure shows that most of the 2,736 studied dependencies exhibit high library residue (i.e., $\bar{x}$=85.7\%, $\mu$=81.5\%, $\sigma$=22.2\%).
An example is the popular \texttt{log4j} logging library \lib{log4j}{1.2.15}, which is an older library, but has a library residue of 98\%.
Finally, Figure \ref{fig:residueVsPLU} shows that the system are more likely to remain with the more popular libraries, with higher peaking libraries exhibiting more library residue. 
Returning to (RQ1):

\begin{hassanbox}
	We conducted an empirical study to understand the extent to which (i) systems use and manage their library dependencies and (ii) library usage trends. To answer (RQ1): \textit{ (i) although system heavily depend on libraries, most systems rarely update their libraries and (ii) systems are less likely migrate their library dependencies, with 81.5\% of systems remaining with a popular older version.}
\end{hassanbox}

\section{Developer Responsiveness to Awareness Mechanisms}
\label{sec:LMT}

\begin{figure*}
	\centering
	\subfigure[LMP for consecutive releases of the \texttt{google-guava} (NR1) library]{\label{fig:guava}
		\includegraphics[width=1\columnwidth]{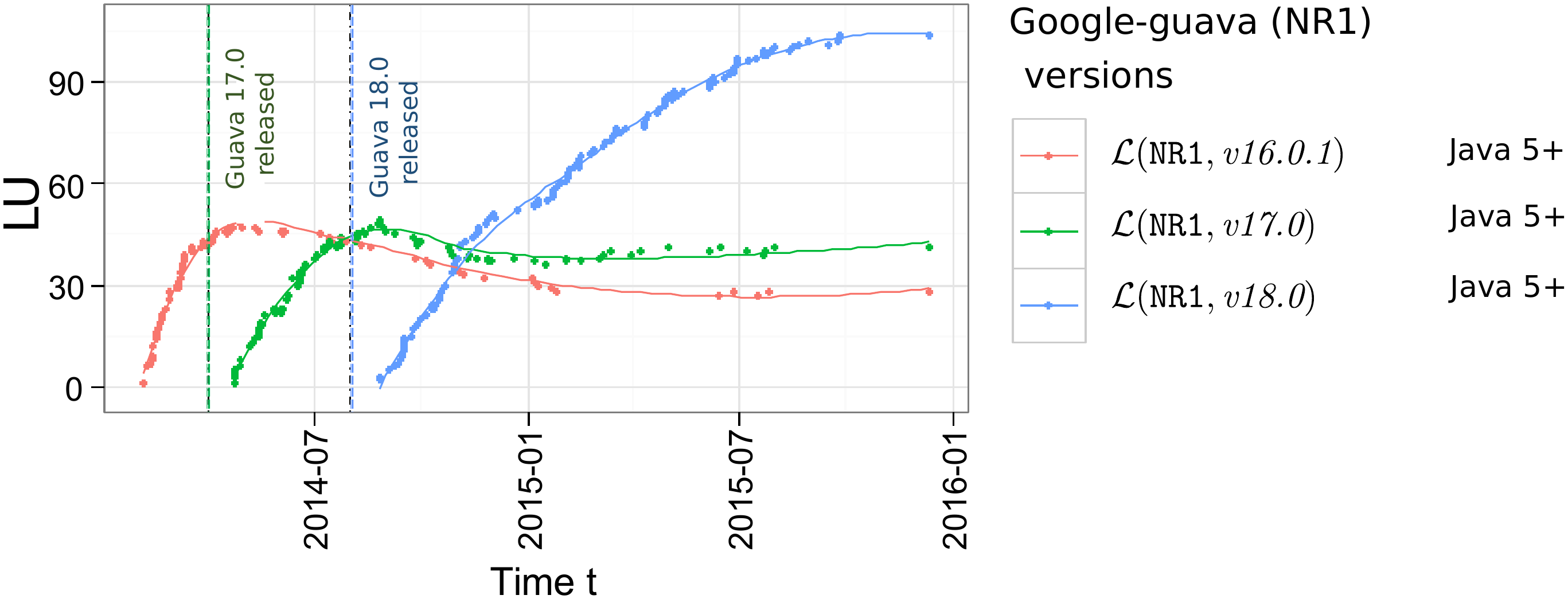}
	}
	\subfigure[LMP for consecutive releases of the \texttt{junit} (NR2) library]{\label{fig:junit}
		\includegraphics[width=1\columnwidth]{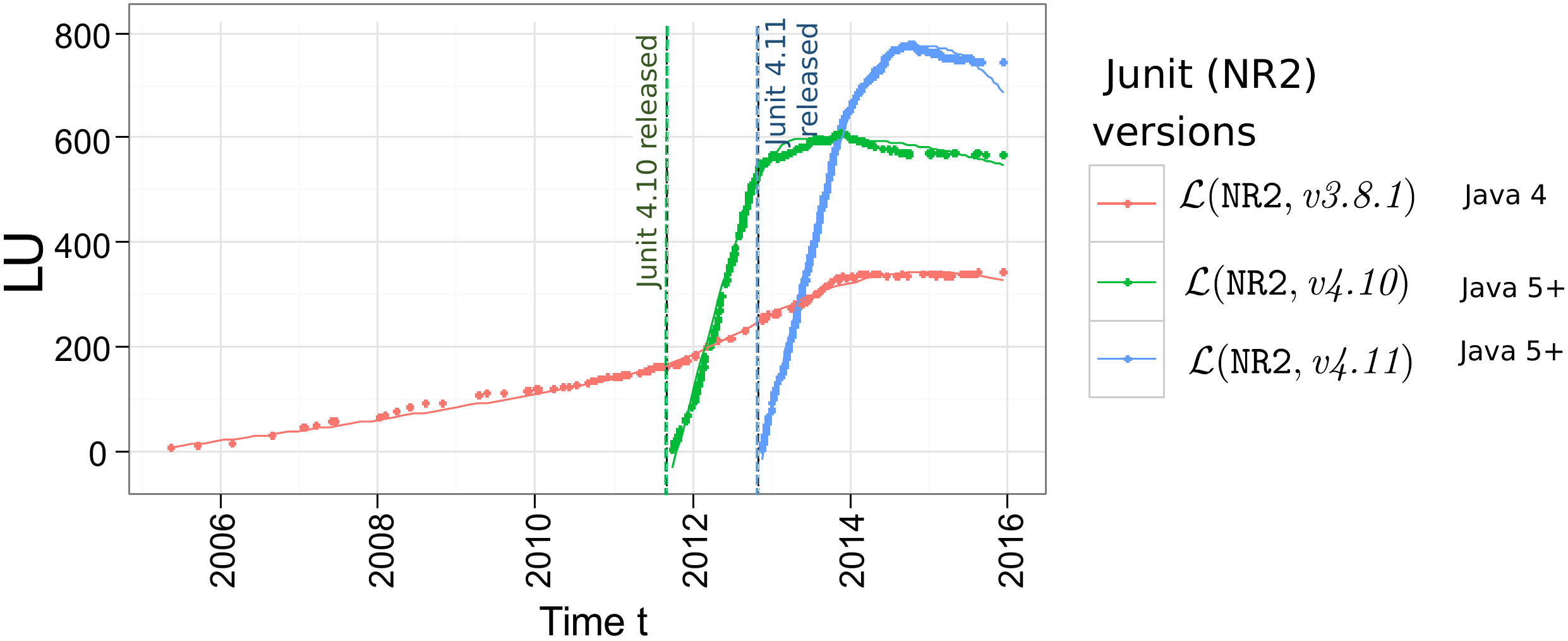}
	}
	\subfigure[LMP for consecutive releases of the \texttt{log4j} (NR3) library]{\label{fig:log4j}
		\includegraphics[width=.9\columnwidth]{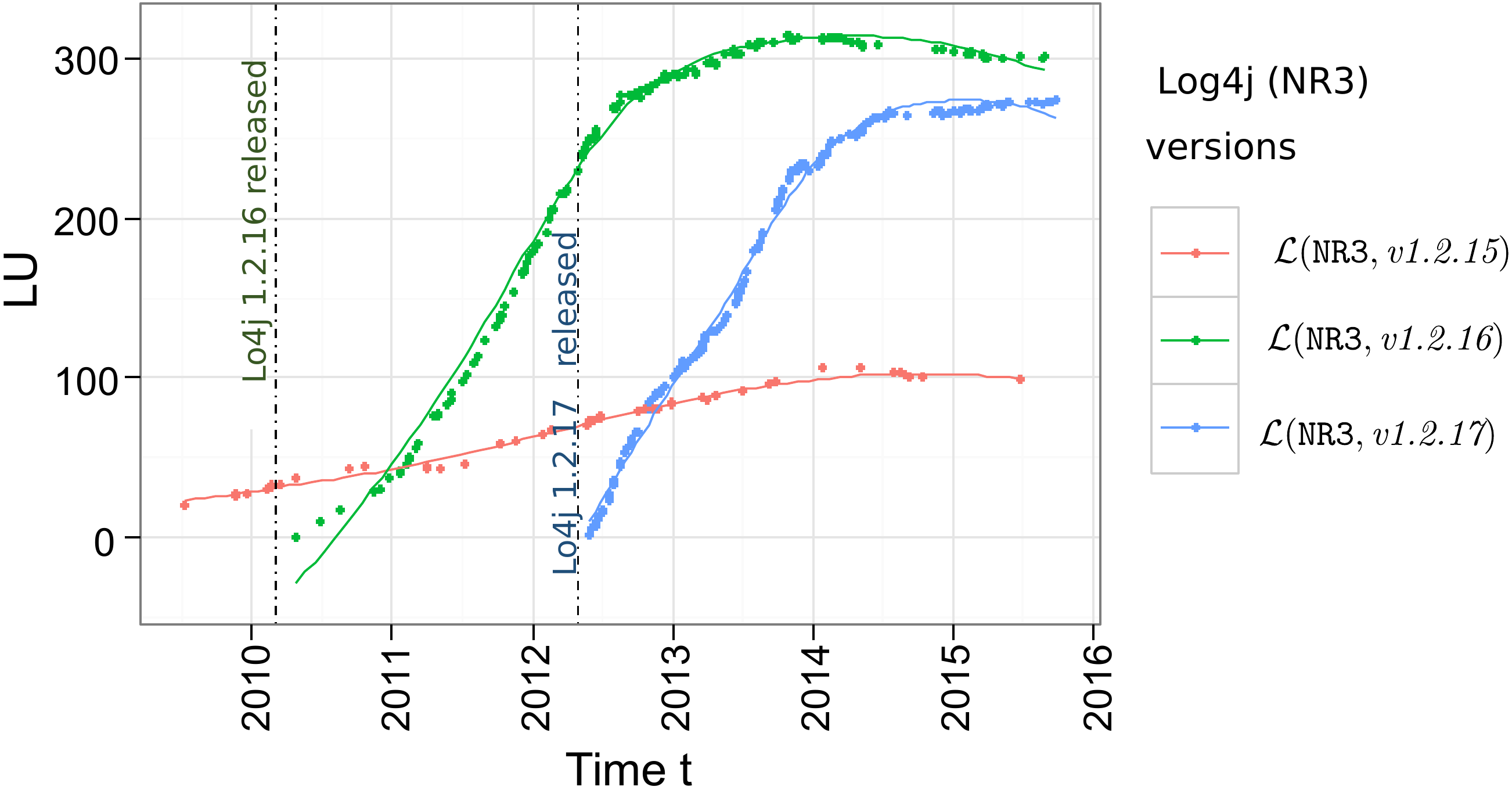}
	}
	\caption{Library Migration Plots (LMP) of three libraries depicting successive library version releases without vulnerability alerts.}
	\label{fig:newVersions}
\end{figure*}

\begin{table*}[b]
	\begin{center}
		\fontsize{7}{10}\selectfont
		\tabcolsep=0.1cm
		\caption{Alias names for our (RQ2) selected case studies.
		}
		\label{tab:vulnerableDataset2}
		\begin{tabular}{lccrcc}
			\hline
			\multirow{1}{*}{\textbf{Alias}}
			& \multirow{1}{*}{\textbf{Awareness Mechanism}}
			& \multirow{1}{*}{\textbf{Library}}
			& \multirow{1}{*}{\textbf{Analyzed versions}}
			\\
			NR1&New Release&{google-guava} &(16.0.1), (17.0), (18.0) \\
			NR2&&{junit}& (3.8.1), (4.10), (4.11)\\
			NR3&&{log4j}& (1.2.15), (1.2.16), (1.2.17)  \\ \hline
			V1&Security Advisory & commons-beautils & (1.9.1), (1.9.2) \\
			V2& & commons-fileupload & (1.2.2), (1.3), (1.3.1) \\
			V3& & commons-httpclient & (3.1), (4.2.2) \\
			V4& & httpcomponents & (4.2.2), (4.2.3), (4.2.5)\\
			V5& & commons-compress &(1.4), (1.4.1) \\ \hline
		\end{tabular}
	\end{center}
\end{table*}

In this section, we present the results for (RQ2) and (RQ3).
To answer (RQ2), \RqTwo, we address in Section \ref{sec:nr} and Section \ref{sec:ad}.
To answer (RQ3), \RqThree, we address in Section \ref{sec:barriers}.
Table \ref{tab:vulnerableDataset2} shows the aliases (\ie~NR1, ..., NR3, V1, ..., V5) used as a reference to each of the case studies.

\subsection{A New Release Announcement}
\label{sec:nr}
Figure \ref{fig:newVersions} depicts our case studies (NR1, NR2, NR3) related to responsiveness of a new release, with (A) consistent and (B) non responsive library migration trends.

\paragraph{(A) Cases of an Active Developer Response to a New Release.}
Figure \ref{fig:guava} shows an example of library that have a consistent library migration trend.
Concretely, the LMP of the \texttt{google-guava} (NR1) \lib{NR1}{16.0.1} and \lib{NR1}{17.0} depicts a consistent pattern of migration with 48 and 49 peak LU.
This pattern is consistent, despite the libraries having a relatively high library residue of 60.4\% and 85\% for all studied versions.

\sloppypar{
We find that the reasons for consistent migration trends are mainly related to the estimated migration effort required to complete the migration process.
Through inspection of the online documentation, we find that migration from \lib{NR1}{16.0.1} to \lib{NR1}{17.0} contains 10 changed packages\footnote{details at \url{http://google.github.io/guava/releases/17.0/api/diffs/}}.
Similarly, migration from  \lib{NR1}{17.0} to \lib{NR1}{18.0} also contained 7 changed packages.
Yet, all three library versions require the same Java 5 environment which indicates no significant changes to the overall architectural design of the library.
From the documentation, we deduce that popular use of \lib{NR1}{18.0} is due to the prolonged period between the next release of \lib{NR1}{19.0}, which is more that a year  after the release of \lib{NR1}{18.0} in December 10, 2015. 
In fact, previous versions had shorter release times, around 2-3 months of \lib{NR1}{16.0.1} in February 03 2014, \lib{NR1}{17.0} in April 22 2014, and \lib{NR1}{18.0} in August 25 2014.
The prolonged released cycles of the library could be related to the relatively higher peak LU of \lib{NR1}{18.0} at 100 LU compared to the lower peaks LU of \lib{NR1}{16.0.1} at 48 LU and 49 LU for the \lib{NR1}{17.0} dependency. }


\paragraph{(B) Cases of a Developer Non Response to a New Release.}
Figure \ref{fig:junit} depicts a developer `no response' reaction to a dependency migration opportunity.
The LMP curve from figure depicts the older popular versions as exhibiting no migration movement (\ie~peak LU= current LU).
Specifically for the \texttt{junit} (NR2) library, the dependency \lib{NR2}{3.8.1} does not follow the typical migration pattern of the \lib{NR2}{4.10} and \lib{NR2}{4.11} dependencies.

Similar to the consistent migration to a new release, we find that the reason for a non response to a migration opportunity is related to the estimated migration effort.
For instance, as shown in Figure \ref{fig:junit}, the newer \texttt{Junit} version 4 series libraries requires a change of platform to Java 5 or higher (\lib{NR2}{4.10} and \lib{NR2}{4.11}), inferring significant changes to the architectural design of the library. 
Intuitively, we see that even though \lib{NR2}{3.8.1} is older, it still maintains its maximum library usage (i.e., current LU and peak LU=342).
This LMP curve pattern is also apparent in the \texttt{log4j} (NR3) library shown in Figure \ref{fig:log4j}, with the \lib{NR3}{1.2.15} dependency being older, but still active library version (\ie~with over 100 current LU). 
We visually observe that as \lib{NR3}{1.2.17} dependency reaches its peak LU the \lib{NR3}{1.2.16} dependency remains more popular, with a higher LU than superseding library release.
This result complements the findings in (RQ1) that popular library dependencies tend to retain most of their users, even if a possible migration to a new release opportunity is available.




\subsection{Security Advisory Disclosure}
\label{sec:ad}
\begin{figure}
	\centering
	\includegraphics[width=1\textwidth]{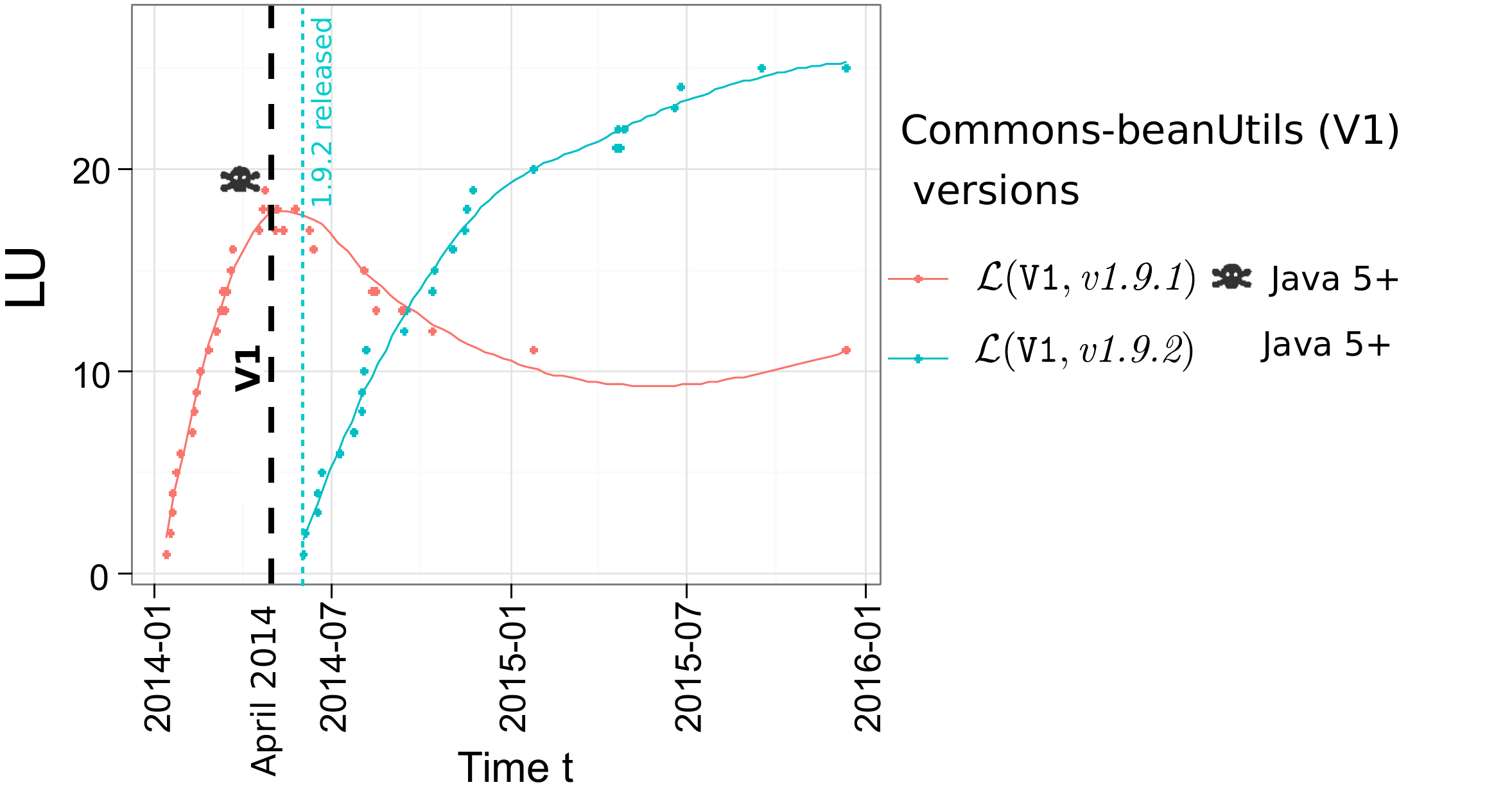}
	\caption{LMP for vulnerability \texttt{V1}, related to the \textsc{commons-beanutils} library dependency versions \texttt{V1}$_{1.9.1}$ and \texttt{V1}$_{1.9.2}$.}
	\label{fig:beanutlisV1}
\end{figure}

\begin{figure*}
	\centering
	\includegraphics[width=1\textwidth]{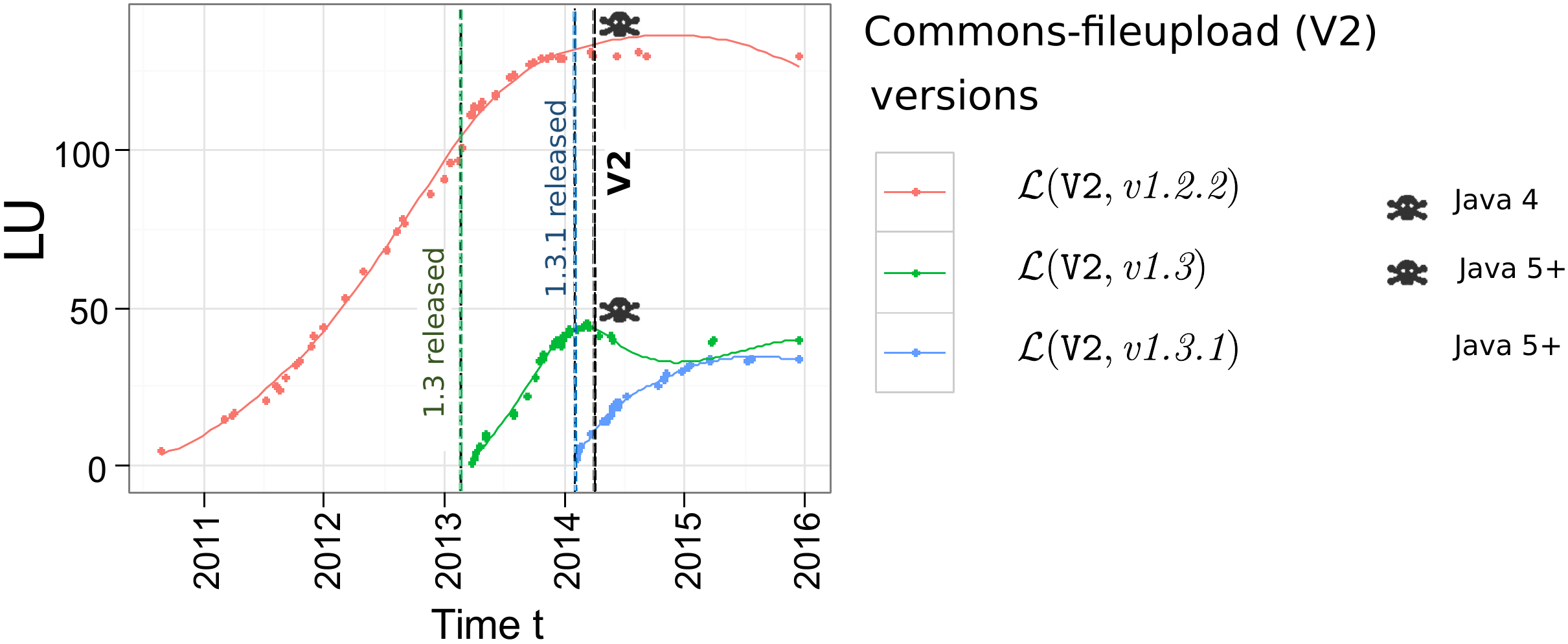}
	\caption{LMP for vulnerability \texttt{V2}, related to the \texttt{commons-fileupload} library dependency versions. }
	\label{fig:fileupload}
\end{figure*}

\begin{figure}
	\centering
	\subfigure[LMP for vulnerability \texttt{V3}, related to the \texttt{commons-httpclient} and the superseeding \texttt{httpcomponents} libraries for vulnerablilty \texttt{V4}. V4 is the fix for the V3 vulnerability that was not fixed for the vulnerable \lib{V4}{4.2.2} dependency.] {\label{fig:httpclient}
		\includegraphics[width=1\columnwidth]{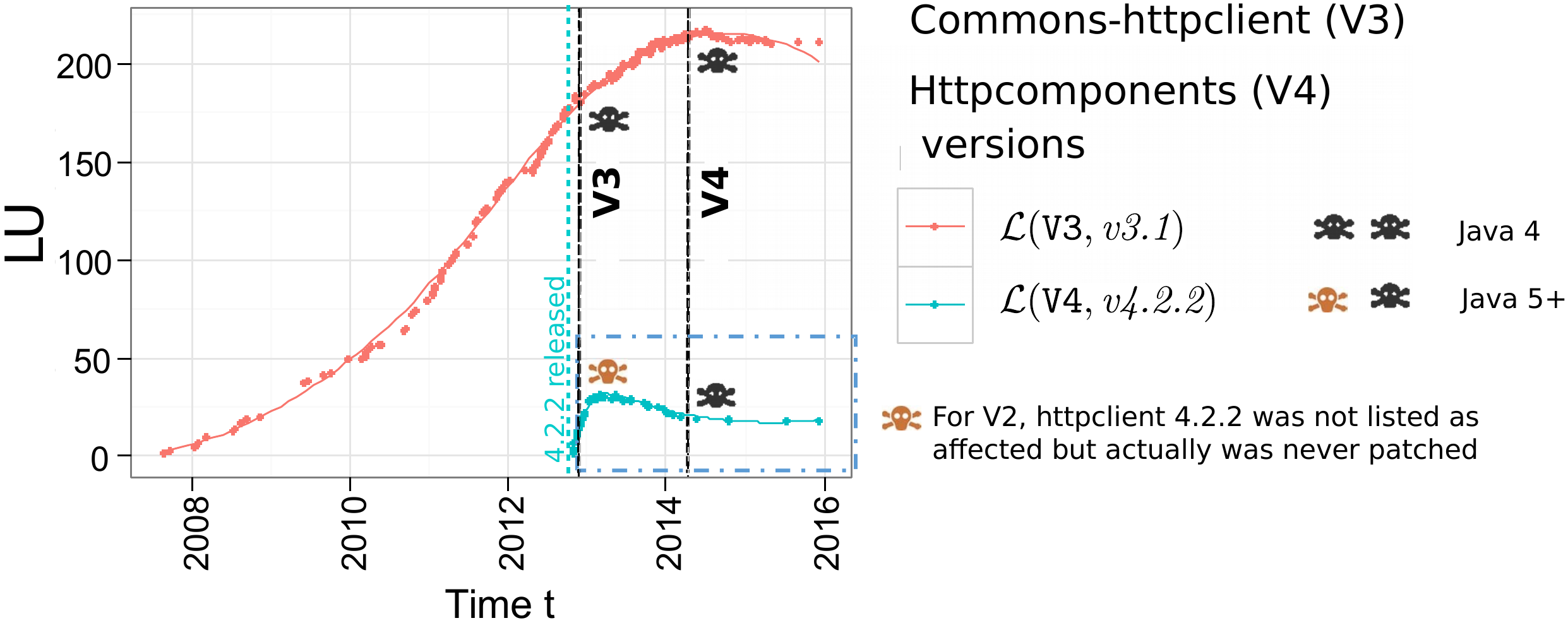}
	}
	\subfigure[LMP for vulnerability \texttt{V4}, related to the \texttt{httpcomponents} library. Note: There seems no effect of V4, possibly because maintainers of the vulnerable \lib{V4}{4.2.2} dependency may have already migrated away to the safer \lib{V4}{4.2.3} and \lib{V4}{4.2.5} dependencies.]{\label{fig:httpcomponents}
		\includegraphics[width=1\columnwidth]{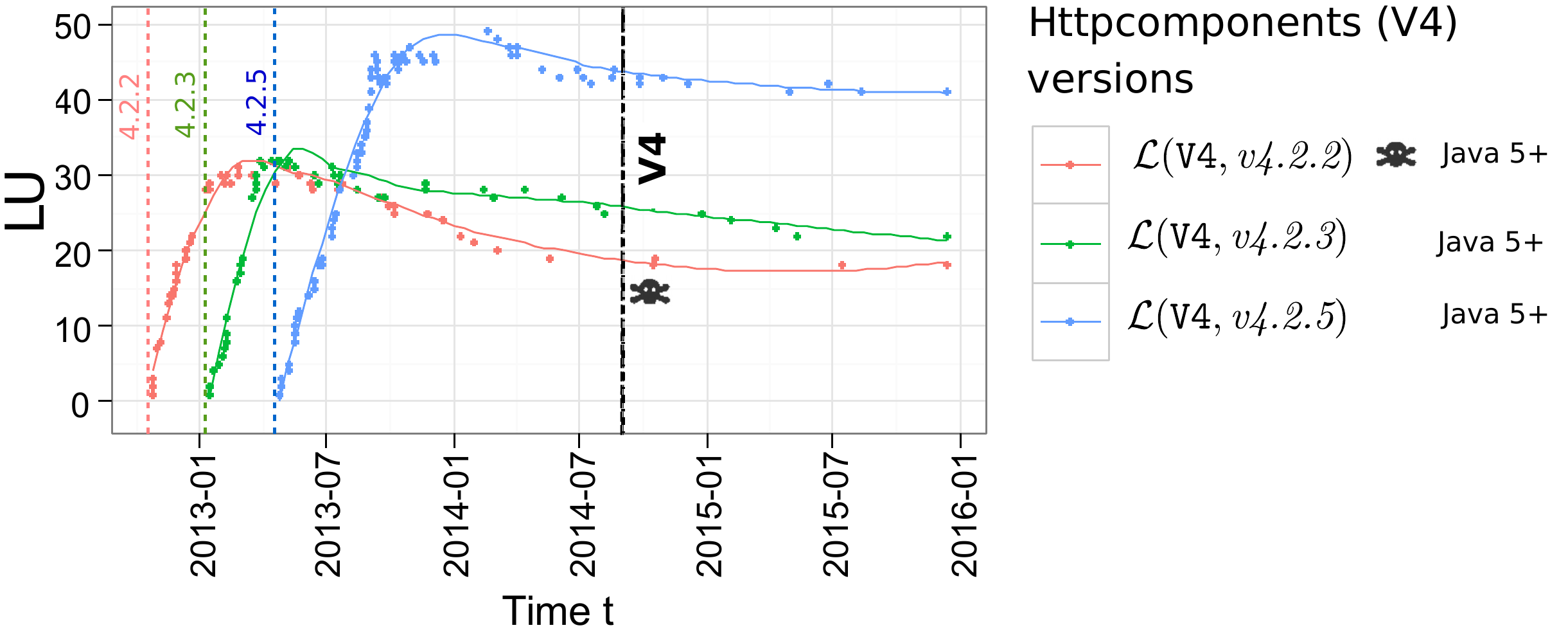}
	}
	\caption{Vulnerability alerts for the \texttt{commons-httpclient} (V3) and related  \texttt{httpcomponents} (V4) library. In detail, Figure \ref{fig:httpcomponents} is a zoomed in look at Figure \ref{fig:httpclient}, which is the vulenerable \lib{V4}{4.2.2} dependency, and safe \lib{V4}{4.2.3} and \lib{V4}{4.2.5} dependencies.}
	\label{fig:V23}
\end{figure}

\begin{figure}
	\centering
	\includegraphics[width=1\textwidth]{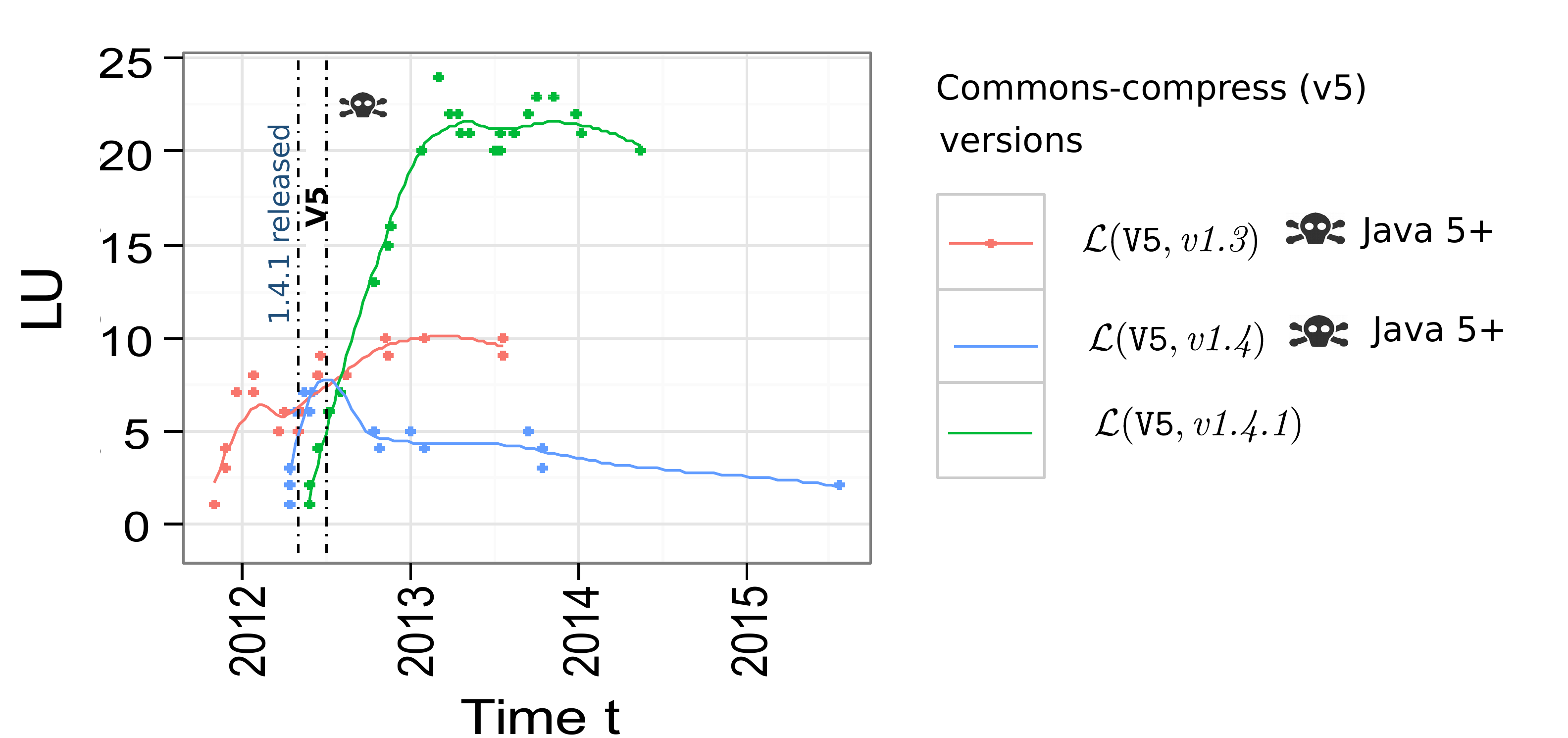}
	\caption{LMP for vulnerability V5, related to the \texttt{commons-compress} library.}
	\label{fig:compress}
\end{figure}

Figures \ref{fig:beanutlisV1}, \ref{fig:fileupload}, \ref{fig:V23} and \ref{fig:compress} all depict the LMP of our case studies related to the responsiveness of affected developers towards a security advisory disclosed to the general public.
In our analysis, we group and discuss the case studies according to (C) an active response, (D) no response and (E) a latent disclosure to a security advisory.

\paragraph{(C) Cases of an Active Developer Response to a Security Advisory Disclosure.}
Figure \ref{fig:beanutlisV1} depicts a typical case of where migration is in response to a vulnerability.
As shown, the LMP curve clearly depicts a peak and decline in the usage after the \texttt{V1} vulnerability security advisory was disclosed to the public.
We conjecture that the timely release of the patched \lib{V1}{1.9.1} dependency shortly after the security advisory was disclosed, provided a migration opportunity for developers.

In contrast to the reported case in \texttt{V1}, Figure \ref{fig:fileupload} depicts a case for \texttt{V2}, where the security for the \texttt{V1} vulnerability security disclosure that affects the \lib{V2}{1.3} dependency does not affect the LMP curve of the older \lib{V2}{1.2.2} dependency.
In detail, the LMP curve is evident by the rise of the \lib{V2}{1.2.2} dependency from 110 LU to 140 LU, during the period in which  the \lib{V2}{1.3.1} dependency is released.
Nevertheless, during this period \lib{V2}{1.3} having also increased from 1 to 48 LU during this period, inferring that during this period, maintainers preferred to adopt the older dependency rather than the newer release.

Meanwhile, the LMP curves in Figure \ref{fig:fileupload} depict how the disclosure of the \texttt{V2} security advisory triggers developers migrating away from both the \lib{V2}{1.2.2} and \lib{V2}{1.3} vulnerable dependencies.
Using the LU trend metrics, we observe that the library residue of these two libraries is still very high with \lib{V2}{1.2.2} has a library residue of 98\%, while \lib{V2}{1.3} reports a 86\% of library residue.
Yet, the LMP curve metrics infer that even though a security advisory is disclosed to the public, many affected developers user systems continue to use an exploitable version.
As in the case of new release announcements, one reason why a developer would not respond to a security advisory is the estimated migration effort required.
In case of the \texttt{V2} vulnerability, inspection of the release logs indicate a relatively high migration effort, as the newer \lib{V2}{1.3} dependency would require an upgrade to Java 5 and higher platform for any system.
Moreover, we conjecture that users of the newer \lib{V2}{1.3} dependency are more likely from developers that have not used prior versions of the affected \texttt{commons-fileupload} library.

\paragraph{(D) Cases of an Incomplete Patch Release in Response to a Security Advisory Disclosure.}

Figure \ref{fig:V23} shows a case where the lack of a replacement dependency may contribute to affected developers showing no response to the disclosure of a security advisory. 
In this case, the initial vulnerability \texttt{V3} is related to the Amazon Flexible Payments Service (FPS), which is a man-in-the-middle attacker to spoof SSL servers via an arbitrary valid certificate.
\texttt{V3} is the original vulnerable \lib{V3}{3.1} dependency that affects users of the \texttt{commons-httpclient} library.
As seen in Figure \ref{fig:httpclient} rising LMP curve, the security advisory does not trigger any migration among its users. 
In fact, there is an increase from 165 to 209 LU after the security alert was disclosed.
Related to V3, V4 is the same man-in-the-middle attach with a \textit{`NOTE: this issue exists because of an incomplete fix for CVE-2012-5783'} in its description\footnote{\url{https://web.nvd.nist.gov/view/vuln/detail?vulnId=CVE-2012-6153}}.

The estimated migration effort and the lack of replacement library are some of the possible reasons why affected maintainers show no response to the security advisory.
This is shown in the case of the \texttt{Httpcomponents} library\footnote{\url{https://hc.apache.org/}}, which is the successor and replacement for \texttt{commons-httpclient} library.
As documented, \texttt{Httpcomponents} is a major upgrade with many architectural design modifications compared to the older \texttt{commons-httpclient} dependency versions.
However, after the first release of the \texttt{Httpcomponents} library, the LMP curve in Figure \ref{fig:httpclient} indicate that  many developers that use this library still actively use the older \texttt{commons-httpclient} dependency versions.
Shown in this figure, after the \textit{V4} security advisory disclosure, did the affected \lib{V4}{3.1} show signs of developers migrating away from the vulnerable dependency.
The LMP curve of the \lib{V4}{3.1} dependency moves from a peak LU of 215 to a decreased 212 LU.

\paragraph{(E) Cases of a Latent Security Advisory Disclosure.}

Figure \ref{fig:httpcomponents} shows a case where affected developers already migrate away from the vulnerable dependency, before the security advisory is disclosed to the public.
In the case of \texttt{httpcomponents} library, its LMP curve indicates that developers who maintain the vulnerable \texttt{V4} \lib{V4}{4.2.2} dependency shows no response to the security advisory disclosure. 
One reason is that a prior migration of the library had already been triggered by the releases of newer versions of \lib{V4}{4.2.3} and \lib{V4}{4.2.5}.
By the time V3 is disclosed, \lib{V4}{4.2.2} is already in decline with a 60\% library residue.

Finally, Figure \ref{fig:compress} depicts a case where the reason for developer responsiveness to a security advisory disclosure cannot be simply explained using the LMP curve. 
In the figure, the LMP curve shows that developers that maintain the vulnerable \texttt{commons-compress} \lib{V5}{1.4} dependency responded to the \textit{V5} security advisory disclosure.
However, this was not the case of all versions of the library.
The LMP curve shows developers that use the older \lib{V5}{1.3} dependency did not show any signs of migrating away from this vulnerable dependency.
In fact, although minor, the LMP curve for the vulnerable \lib{V5}{1.4} dependency rises in LU after the security advisory was disclosed to the public.
Returning back to (RQ2):
%

\begin{hassanbox}
	We conducted an empirical study to understand developer responsiveness to (i) a new release and (ii) a security advisory disclosure.
	To answer RQ2: \textit{we find that for a new release of a popular library (i) there exist patterns of consistent migration and patterns where an older popular library version is still preferred.
	For a security advisory disclosure we find cases of developer (ii) non responsiveness to security advisory disclosure, which  is sometimes due to an incomplete patch or a latent security advisory.
	We find developers are less likely to update a library that requires more migration effort and vice-versa. }
\end{hassanbox}

\subsection{Developer Feedback on Updating a Vulnerable Dependency}
\label{sec:barriers}

\begin{table*}
	\begin{center}
		\fontsize{8}{10}\selectfont
		\tabcolsep=0.1cm
		\caption{Summary of the survey collected from projects with a known security vulnerability.
		}
		\label{tab:surveyInfo}
		\begin{tabular}{lcc|cccc}
			\hline
			\multirow{1}{*}{\textbf{Alias}}
			& \multirow{1}{*}{\textbf{\# Listed}}
			& \multirow{1}{*}{\textbf{\# Contactable}}
			& \multirow{1}{*}{\textbf{\# Feedback}}
			& \multirow{1}{*}{\textbf{Unaware}}
			& \multirow{1}{*}{\textbf{Updated}}
			\\\hline
			V1&42&23&5&4&4\\
			V2&40&26&6&6&5\\
			V3&20&5&1&0&1\\
			V4&10&7&3&1&0\\
			V5&8&3&1&0&1\\\hline
			\textbf{Totals}&\textbf{120}&\textbf{64}&\textbf{16}&\textbf{11}&\textbf{11}\\
		\end{tabular}
	\end{center}
\end{table*}

Table \ref{tab:surveyInfo} shows a summary of affected projects that show negligence in responding to any of the five (\ie~V1, V2, V3, V4, V5) security advisories analyzed in (RQ2). 
From the 120 projects that were detected by the LMP curve, we found 64 of the projects provided a feedback mechanism such as a mailing list or issue management system.
As shown, out of the 64 projects, 16 projects (25\%) of the projects provided us feedback.
In this section, we discuss the results pertaining to: (F) developer awareness on the vulnerability affecting their projects and (G) developer opinion regarding the practice of updating dependencies.

\paragraph{(F) Developer Awareness of Vulnerabilities in their Projects.}
Table \ref{tab:surveyInfo} shows that many of the affected projects were unaware of the vulnerability to their software. 
Through the feedback, we find that out of the 16 responses, 11 (69\%) immediately thanked us for the notification and proceeded to update their dependencies to the safer dependency versions.

\paragraph{(G) Developer Opinion on Library Updates.}
Developers cite the threat of the impact of the exposure as well as the function of the dependency as a factor to influence the decision in responding to a security advisory.
A developer from a project responded that \textit{`our project has been inactive and production has been halted for indefinite time.'} while another.
Developers from another two projects noted that the vulnerable dependency did not have a critical effect on the project:
 \begin{quote}
 	\small{ \textit{`I knew about it because I happen to work on another project where we had to fix this very problem, but I didn't connect two dots.
 	In this case, it's a test dependency, so the vulnerability doesn't really apply.'}} and that \small{  \textit{`It's only a test scoped dependency which means that it's not a transitive dependency for users of XXX so there is no harm done.
 	XXX has no external compile scoped dependencies thus there is no real need to update dependencies.'}}
 \end{quote}

 \noindent Finally, the remaining two projects stated that the update was unnecessary as the affected component had little impact on the project objectives or part of their responsibilities.
 A developer from the first project stated that \textit{`When it comes to this specific vulnerability org.apache.commons.httpclient is only used by XXX by the automatic update, and there's no SSL or encryption involved.'} while another developer from the latter project deferred the responsibility to another project \textit{` We don't maintain XXX, but have passed this on to XXX. Not our decision, this is a slightly revised fork.'}

Similar to the results for (RQ2), all developers from the 16 projects cite the required migration effort as an influencing factor to whether or not a vulnerable library should be updated. 
For instance, one developer discusses how security updates are not a priority, as they do not align with the goals and objectives of their software customers:

 \begin{quote}
 	\small{ \textit{`I subscribed to the CVE RSS recently and I don't check it regularly, so even
 			if I might have heard of the current vulnerability, I simply forgot to address
 			it. We also had some emergencies recently (developing features for our customers),
 			that makes the security issues less prio than releasing the ordered features :-/
	 		} ... \textit{ Anyway, our security approach is far from perfect, I am aware of it, and I'm
	 		willing to improve this, but sometimes it is difficult to explain our customers
	 		that it is a main point to consider in the development process.'}}
 \end{quote}

\noindent In other developer feedback, we find developers perceive the practice of updating their dependencies as added effort and responsibilities that should be performed in their \textit{`spare time'}.
Moreover, developers suggest that availability of the manpower and assigned role as factors for deciding whether or not to migrate the dependency:

  \begin{itemize}
  	\item \small{ \textit{`Just that it's not very easy to keep track of it. As there's no downside in upgrading in this case, we would have done so if for example there's a build time warning about such dependency....	Thank you very much for bringing this to our attention!  We don't maintain XXX, but have passed this on to XXX. Not our decision, this is a slightly revised fork.'}}

 \item
 	\small{  \textit{`I suppose we weren't aware of this issue.'} and that they have issues with the \textit{`amount of people currently working on the project with spare time to verify it works correctly with new version'.}}

 \item
 	\small{  \textit{`I can't answer for the group, but generally there are so many security vulnerabilities that it's a full time job just to keep up with them all. In most cases they don't apply.'}}

  \item
  	\small{\textit{`As mentioned above, we are no longer maintaining this particular version.
  			Yes, but only "decentralized and informal one": whoever introduced the dependency is supposed to keep track of it and update it in the part(s) he is maintaining.'}}
  \end{itemize}

\noindent Finally, returning to (RQ3):


\begin{hassanbox}
	We contacted developers to understand developer (i) awareness and (G) opinions regarding the practice of updating their library dependencies.
	To answer RQ3: \textit{We find that 69\% of developers were unaware of their vulnerable dependencies and proceeded to immediately migrate to a safer dependency.
	Developers evaluate the decision whether or not to update its dependencies based on project specific priorities. 
	Developers cite migration as a practice that requires extra migration effort and added responsibility.}
\end{hassanbox}

\section{Discussion}
\label{sec:dis}
In this section, we discuss the implications of our results and the validity of our study.

\subsection{Implications of Results}
To understand implications of the study, we first speculate some factors that influence the decision for a developer to migrate a dependency. 
Then, we show how our work is relevant in the context of existing literature on updating third-party libraries.
Results for (RQ1) show that contemporary projects heavily use libraries, often forming complex inter-dependencies within the project.
Implied responses for (RQ3), the complex nature of these inter-dependencies (colloquially termed as \textit{`dependency hell'}) contributes to the migration effort needed to update a dependency.
When addressing (RQ2) and (RQ3), we find that updating dependencies is a cost--benefit decision between estimating the amount of migration effort needed and evaluating the benefits of adopting a replacement dependency.
Finally, we speculate that an overwhelm of staff responsibilities and a lack of motivation may play a role in the decision whether to update or not a dependency.
Developer feedback also indicate that library migration as being \textit{low priority} and  \textit{`extra work to be done in spare time'}.

Our findings are consistent with prior work that studied library updates at the API level.
Indeed, at the API level of abstraction, migration effort defined in terms of the different API calls between a library and client system.
In our study, we consider at a higher level of abstraction, with overall migration effort defined as additional rework, testing and compatibility with other inter-dependencies when updating a dependency.
Our work complements related work in several ways.
For instance, \cite{Robbes:2012} states that: \textit{``A minority of reactions to API changes can remain undiscovered long after the original change is introduced"}.
Our results does concur with this work, yet, with developers claiming to be unaware of opportunities to migrate their dependencies.
A study by \cite{Bavota:2015} comprehensively investigates API and library updates within the Apache ecosystem, with the goal of understanding what different product and process factors may lead to developers updating their libraries.
The study shows that at the API level, developers tends to upgrade a dependency when substantial changes such as bug-fixing activities are included in the replacement dependency.
Our work finds that developers regard library maintenance as extra work.
Hence, we speculate that developers may update a dependency if (i) they are aware of a migration opportunity, (ii) they have the time and responsibility to perform the migration effort, and (iii) they feel the update aligns with the goals and objectives of their project.
Furthermore, \cite{McDonnell2013} states that: \textit{``Android APIs are evolving fast and client adoption is not catching up with the pace of API evolution"}.
This result is further strengthened by the findings of \cite{hora:2015}, who says that: \textit{``...53\% of the analyzed API changes caused reaction in 5\% of the systems and affected 4.7\% of developers".}
To complement this findings of \cite{Bogart:SCGSE15}, we reveal that developers are unaware of awareness mechanisms, and non-technical organization factors play an important role in the developer decision to update.

\subsection{Threats to Validity}
\label{sec:threat}
We now present construct, internal and external threats to our study.

\paragraph{Construct Validity - }
refers to the concern between the theory and achieved results of the study.
We find three threats that relate to the tools and mechanisms used to obtain our results.
The first is source of our datasets.
In reality, there are other forms of awareness mechanism such as social media alerts or the word--of--mouth medium, to raise developer awareness to a migration opportunity.
However, we believe that new releases and security advisories are more traditional and recognized forms of announcements.
For future work, we could also investigate other forms of awareness mechanisms that lead to a library migration.
The second threat is the related to the tools used to extract our dependency migration information.
In this study, we use the configuration file of the Maven dependency manager to assume the third-party dependencies.
There may be cases were a third-party library is not declared in such configuration, and is instead embedded manually into the system.
Our tool \textit{PomWalker} cannot capture these dependencies as it specifically detects documented dependency declarations (\ie{} implicit version references and managed dependencies).
Our method also does not count dependencies that have copied the source code of the library into their own source code.
However, since our collected dependencies under-estimates of actual reuse, we do not believe it to affect our main result, thus mitigating this threat.
We also assume that explicit stated versioning are used by projects that are more likely to manage their dependencies.
The final threat is the selection criteria used to select the case studies for (RQ2).
The threat is that the criteria was performed manually (matching CVE to libraries) which could be error-prone and might have missed other case studies.
However, we believe that our systematic research methods ensure a quality dataset and case studies that validate our drawn conclusions.

\paragraph{Internal Validity - }
refers to the concerns that are internal to the study.
In this study, we found two main internal threats that could affect our results.
First is the accuracy of our dataset and generalization of our results to represent the real world.
This has an impact to both (RQ1) and (RQ2).
For instance, a dataset containing obsolete projects or inactive forks of systems would cause false positives on the Library Migration Plot (LMP) trend curve.
To mitigate this, we took particular care to filter out projects that were not regularly maintained by its developers.
The second threat is related to the research method and actual response rate of the developers for RQ3.
We understand that these little responses cannot be a true representation of all developers.
However, we believe that the response rate of 25\% from the smaller set of contactable projects is indeed adequate when targeting a specific interest group.

\paragraph{External Validity - }
refers to the generalization concerns of the study results.
We have two main threats to the results of the study.
First is the conclusions of the case studies to generalize the trend patterns for all Java projects.
Due to dataset quality pre-processing, we analyze projects that do use third-party libraries, hence, results may not be applicable to the other types of Java projects that do not use a dependency management tool.
Also, there is a threat to the case studies not being representative of all projects.
For new releases, we found that the three libraries depicted the typical pattern of either system users consistently migrating or where an older library version is the popular version. There may be other interesting patterns, but theoretically for new releases, this are the typical two patterns for popular libraries.
The second threat is the generalization to other library ecosystems such as JavaScript npm, Ruby RubyGems.
We are careful to restrict these findings to Java projects as other ecosystems may depict a different set of library migration patterns and tendencies.
This would be an interesting future avenues for research.
We envision that different lessons learnt from other ecosystems in terms of responsiveness to library updates may provide insights on how to encourage library maintenance within the Maven ecosystem.


\vspace{-0.2cm}
\section{Related Work}
\label{sec:related}
Complementary to the related work of \cite{Robbes:2012}, \cite{hora:2015}, \cite{McDonnell2013} and \cite{Bavota:2015} already presented in the paper, there has been other work that have studies library migrations, both at the API and library component level.
In this section, cover the body of API literature on library popularity, API library migrations and studies on software ecosystems.

\paragraph{API Library Updates -}
\cite{Teyton:2014} studied library migrations of Java open source libraries from a set of client with a focus on library migration patterns.
The main result of that study was that recommendations of libraries could be inferred from the analysis of the migration trends.
In this work, we have a different motivation to how much migration occurs and especially in relation to vulnerabilities.
Another work was by \cite{Xia2013}, that studied the reuse of out-dated project written in the c-based programming language.
\cite{Kabinna:2016} and colleagues especially focused on the migration of specific logging libraries and not related to vulnerabilities.

Recently, large-scale empirical studies have been conducted on library updates.
\cite{RaemaekersICSM} performed several empirical studies on the Maven repositories about the relation between usage popularity and system properties such as size, stability and encapsulation.
\cite{Raemaekers2014} also studied the relationship between semantic versioning and breakages.
Other related empirical studies were conducted by \cite{Jezek2015} and \cite{Cox:2015}. They studied in-depth how libraries that reside in the Maven Central super repository evolve.
The motivation of our work differs from those work, as we are more focused on the migration process itself and its triggers rather than the migration effort needed to migrate a dependency.

\paragraph{Library Usage as popularity measures - }
The LU metrics and the LMPs are forms of popularity measure by the crowd.
Popularity is not a new concept, with several research on usage trends of libraries.
There has been work that have analyzed different dimensions on library usage by clients.
For example, work such as \cite{DeRoover2013} exploited library usage at the API level to understand popularity and usage patterns of clients.
Similarly, they also looked at both the system and library dimensions of API usage for the Qualitas dataset of projects.
The main differences to our work is that although overlapping, De Roover and colleagues analyzed at the API level, where we look at the higher abstraction of the library level.
Moreover, instead of a simple popularity count, we define a model and metrics to quantify different metrics of LU.

Much like the LMP, related studies have used library usage visually to measure stability \citep{McDonnell2013} or popularity \citep{Mileva:2009}.
In this context, our previous work \citep{2014VISSOFTKula}, among work leveraged popularity to recommend when libraries are deemed safe to use by the masses.
Popularity has also been leveraged in IDEs.
For instance, \cite{EisenbergEtAl10a} improve navigation through a library's  structure using the popularity of its elements to scale their depiction.
Recently, \cite{hora2015apiwave} introduced apiwave in visualizations to show popularity trends at the API level.

\paragraph{Library migration support - }
There has been much research related to the transformation of client code to support library migration, particularly pertaining to the migration effort required.
Work by \cite{Chow:1996} and \cite{Balaban:2005} use a change specification language.
\cite{Wu:2015APIEvoluion} showed in an empirical study that imperfect change rules can be used by developers upgrading their code, especially when documentation is lacking.
There is work that provide the client automatic tool support to accommodate changes made to a library.
For instance, SemDiff by \cite{Dagenais:2009} recommends replacements for framework methods that were accessed by clients.
Other similar tools were proposed by \cite{Xing2007} and \cite{Schafer:2008}.
In this work, we propose to view the migration from a higher level of abstraction at the library component level.

These tools also do not consider the other aspects of the migration process.
Closely related to our work, \cite{Plate:ICSME2015} states that impact assessment, migration effort, and the customer are issues faced by the pragmatic developers wanting to update their vulnerable libraries.
This study shows that these are indeed some of the reasons why maintainers are not updating, even in cases where it exposes the software to outside malicious attacks.

Other work on reuse support is through code analysis.
This area of work considers code clone detection techniques by \cite{KamiyaTSE2002} to support which library version is most appropriate candidate for migration.
\cite{Godfrey2005} proposed origin analysis to recover context of code changes.
Our previous work by \cite{Kawamitsu2014} tracked how code is reused across different code repositories.
Also, work such as \cite{Cossette2012} depict the complexities of the migration effort needed for library changes and transformations at the API level.

\paragraph{Software systems as ecosystems.}
\cite{lungu2008} best termed ecosystems as a `collection of software projects which are developed and evolved together in the same environment'.
The discussed work of \cite{Robbes:2012}, \cite{McDonnell2013} and \cite{Bavota:2015} involved the analysis of API usage within a software ecosystem.
Related, \cite{Wu2015APIUsages} explored the API changes and usages on Apache and Eclipse ecosystems.
In this work, we also look at the Maven Java ecosystem of libraries, however, our clients are indeed from `wild' real-world projects that reside in the much more diverse GitHub repository of repositories.
More recent work has been by \cite{Wittern:MSR2016}, who studied dynamics of the npm JavaScript library ecosystem.

\cite{Mens:ECOS} perform ecological studies of the R CRAN open source software ecosystems.
\cite{2013:wea:haenni} performed a survey to identify the information that developers lack to make decisions about the selection, adoption and co-evolution of upstream and downstream projects in a software ecosystem.
Similar works were performed by \cite{2013CSMRGerman} for the R software ecosystem.
The external library dependencies could be considered as part of the ecosystem.
Therefore, the larger ecosystem of library dependencies may also trigger migrations.
However, in this study, we focused on the trigger effect of vulnerabilities and updates within the same library.





\section{Conclusion}
\label{sec:conclude}
Many software projects today advocate the use of third-party libraries because of its many benefits for software developers.
However, results of this study show that updates of third-party library dependencies are not regularly practiced, especially to fix vulnerabilities that exploit a system to attackers.
Surprisingly, we found that 81.5\% of our studied systems still remain with an outdated dependency.
The study shows many factors that influence the decision whether or not to update a library.
Migration effort such as rework required prepare a system to work on a new platform (\ie~Java 4 to Java 5) and address the API changes plays an important role in the update decision.
Developer awareness also influences the migration process and they do not prioritize updates by questioning the migration cost, citing it as added responsibility and effort to be performed in their \textit{`spare time'}.
We speculate that other issues include an overwhelm of staff responsibilities and a lack of motivation play a role in the decision whether to update or not.

The study provides motivation for our community develop strategies to improve a developer personal perception of third-party updates, especially in cases when effort must be allocated to mitigate a severe vulnerability risk.
Visual aids such as the Library Migration Plots (LMP) provide a rich visual analysis, which could prove useful awareness and motivation for developers quickly update.
For future work, we plan to further explore the developers perception on migration effort.
Specifically, we would like to better understand the responsibilities of updating when using a third-party library dependency.

\begin{acknowledgements}
This work is supported by JSPS KANENHI (Grant Numbers JP25220003 and JP26280021) and the ``Osaka University Program for Promoting International Joint Research."
\end{acknowledgements}


\bibliographystyle{plainnat}
\bibliography{sigproc}

\begin{thebibliography}{39}
\providecommand{\natexlab}[1]{#1}
\providecommand{\url}[1]{\texttt{#1}}
\expandafter\ifx\csname urlstyle\endcsname\relax
  \providecommand{\doi}[1]{doi: #1}\else
  \providecommand{\doi}{doi: \begingroup \urlstyle{rm}\Url}\fi

\bibitem[Balaban et~al.(2005)Balaban, Tip, and Fuhrer]{Balaban:2005}
Ittai Balaban, Frank Tip, and Robert Fuhrer.
\newblock Refactoring support for class library migration.
\newblock In \emph{Proceedings of the 20th Annual ACM SIGPLAN Conference on
  Object-oriented Programming, Systems, Languages, and Applications}, OOPSLA
  '05, pages 265--279, New York, NY, USA, 2005. ACM.
\newblock ISBN 1-59593-031-0.

\bibitem[Bavota et~al.(2015)Bavota, Canfora, Di~Penta, Oliveto, and
  Panichella]{Bavota:2015}
Gabriele Bavota, Gerardo Canfora, Massimiliano Di~Penta, Rocco Oliveto, and
  Sebastiano Panichella.
\newblock How the apache community upgrades dependencies: An evolutionary
  study.
\newblock \emph{Empirical Softw. Eng.}, 20\penalty0 (5):\penalty0 1275--1317,
  October 2015.
\newblock ISSN 1382-3256.

\bibitem[Bogart et~al.(2015)Bogart, K{\"a}stner, and Herbsleb]{Bogart:SCGSE15}
Christopher Bogart, Christian K{\"a}stner, and James Herbsleb.
\newblock When it breaks, it breaks: How ecosystem developers reason about the
  stability of dependencies.
\newblock In \emph{Proceedings of the ASE Workshop on Software Support for
  Collaborative and Global Software Engineering (SCGSE)}, 11 2015.

\bibitem[Chow and Notkin(1996)]{Chow:1996}
Kingsum Chow and David Notkin.
\newblock Semi-automatic update of applications in response to library changes.
\newblock In \emph{Proceedings of the 1996 International Conference on Software
  Maintenance}, ICSM '96, Washington, DC, USA, 1996. IEEE Computer Society.

\bibitem[Cossette and Walker(2012)]{Cossette2012}
Bradley~E. Cossette and Robert~J. Walker.
\newblock {Seeking the ground truth}.
\newblock \emph{Proc. of the ACM SIGSOFT Intrn. Symp. on the Foundations of
  Software Engineering - FSE '12}, 2012.

\bibitem[Cox et~al.(2015)Cox, Bouwers, van Eekelen, and Visser]{Cox:2015}
Joel Cox, Eric Bouwers, Marko van Eekelen, and Joost Visser.
\newblock Measuring dependency freshness in software systems.
\newblock In \emph{Software Engineering (ICSE), 2015 IEEE/ACM 37th IEEE
  International Conference on}, volume~2, pages 109--118, May 2015.

\bibitem[Dagenais and Robillard(2009)]{Dagenais:2009}
Barthelemy Dagenais and Martin~P. Robillard.
\newblock Semdiff: Analysis and recommendation support for api evolution.
\newblock In \emph{Proceedings of the 31st International Conference on Software
  Engineering}, ICSE '09, pages 599--602, Washington, DC, USA, 2009. IEEE
  Computer Society.
\newblock ISBN 978-1-4244-3453-4.

\bibitem[{De Roover} et~al.(2013){De Roover}, Lammel, and Pek]{DeRoover2013}
Coen {De Roover}, Ralf Lammel, and Ekaterina Pek.
\newblock {Multi-dimensional exploration of API usage}.
\newblock \emph{IEEE International Conference on Program Comprehension}, pages
  152--161, 2013.

\bibitem[Edgell and Noon(1984)]{Edgell84}
S.~Edgell and S.~Noon.
\newblock Effect of violation of normality on the t test of the correlation
  coefficient.
\newblock In \emph{Psychological Bulletin}, pages 576--583, 1984.

\bibitem[Eisenberg et~al.(2010)Eisenberg, Stylos, Faulring, and
  Myers]{EisenbergEtAl10a}
Daniel~S. Eisenberg, Jeffrey Stylos, Andrew Faulring, and Brad~A. Myers.
\newblock Using association metrics to help users navigate {API} documentation.
\newblock In \emph{VL/HCC2010}, pages 23--30, 2010.

\bibitem[German et~al.(2013)German, Adams, and Hassan]{2013CSMRGerman}
Daniel~M. German, Bram Adams, and Ahmed~E. Hassan.
\newblock The evolution of the r software ecosystem.
\newblock \emph{Proc. of European Conf. on Soft. Main. and Reeng. (CSMR2013)},
  pages 243--252, 2013.

\bibitem[Godfrey and Zou(2005)]{Godfrey2005}
M.W. Godfrey and L.~Zou.
\newblock {Using origin analysis to detect merging and splitting of source code
  entities}.
\newblock \emph{IEEE Transactions on Software Engineering}, 31\penalty0
  (2):\penalty0 166--181, February 2005.

\bibitem[Haenni et~al.(2013)Haenni, Lungu, Schwarz, and
  Nierstrasz]{2013:wea:haenni}
Nicole Haenni, Mircea Lungu, Niko Schwarz, and Oscar Nierstrasz.
\newblock Categorizing developer information needs in software ecosystems.
\newblock In \emph{Proc. of Int. Work. on Soft. Eco. Arch. (WEA13)}, pages
  1--5, 2013.

\bibitem[Hora and Valente(2015)]{hora2015apiwave}
Andr{\'e} Hora and Marco~Tulio Valente.
\newblock apiwave: Keeping track of api popularity and migration.
\newblock In \emph{International Conference on Software Maintenance and
  Evolution}, 2015.

\bibitem[Hora et~al.(2015)Hora, Robbes, Anquetil, Etien, Ducasse, and
  Valente]{hora:2015}
Andre Hora, Romain Robbes, Nicolas Anquetil, Anne Etien, Stephane Ducasse, and
  Marco~Tulio Valente.
\newblock How do developers react to api evolution? the pharo ecosystem case.
\newblock In \emph{Proceedings of the 2015 IEEE International Conference on
  Software Maintenance and Evolution (ICSME)}, ICSME '15, pages 251--260,
  Washington, DC, USA, 2015. IEEE Computer Society.
\newblock ISBN 978-1-4673-7532-0.
\newblock \doi{10.1109/ICSM.2015.7332471}.

\bibitem[Jezek et~al.(2015)Jezek, Dietrich, and Brada]{Jezek2015}
Kamil Jezek, Jens Dietrich, and Premek Brada.
\newblock {How Java APIs break - An empirical study}.
\newblock \emph{Information and Software Technology}, pages 129--146, 2015.
\newblock ISSN 09505849.
\newblock \doi{10.1016/j.infsof.2015.02.014}.

\bibitem[Kabinna et~al.(2016)Kabinna, Bezemer, Shang, and Hassan]{Kabinna:2016}
Suhas Kabinna, Cor-Paul Bezemer, Weiyi Shang, and Ahmed~E. Hassan.
\newblock Logging library migrations: A case study for the apache software
  foundation projects.
\newblock In \emph{Proceedings of the 13th International Workshop on Mining
  Software Repositories}, MSR '16, pages 154--164, New York, NY, USA, 2016.

\bibitem[Kamiya et~al.(2002)Kamiya, Kusumoto, and Inoue]{KamiyaTSE2002}
T.~Kamiya, S.~Kusumoto, and K.~Inoue.
\newblock {CCFinder:} a multilinguistic token-based code clone detection system
  for large scale source code.
\newblock \emph{IEEE Transactions on Software Engineering}, 28\penalty0
  (7):\penalty0 654--670, 2002.
\newblock ISSN 0098-5589.
\newblock \doi{10.1109/TSE.2002.1019480}.

\bibitem[Kawamitsu et~al.(2014)Kawamitsu, Ishio, Kanda, Kula, Roover, and
  Inoue]{Kawamitsu2014}
Naohiro Kawamitsu, Takashi Ishio, Tetsuya Kanda, Raula~Gaikovina Kula, Coen~De
  Roover, and Katsuro Inoue.
\newblock Identifying source code reuse across repositories using lcs-based
  source code similarity.
\newblock In \emph{Proc. of SCAM}, 2014.

\bibitem[Kula et~al.(2014)Kula, Roover, German, Ishio, and
  Inoue]{2014VISSOFTKula}
Raula~G. Kula, Coen~D. Roover, Daniel~M. German, Takashi Ishio, and Katsuro
  Inoue.
\newblock Visualizing the evolution of systems and their library dependencies.
\newblock In \emph{Proc. of IEEE Work. Conf. on Soft. Viz. (VISSOFT)}, ICSME
  '15, 2014.

\bibitem[Kula et~al.(2015)Kula, German, Ishio, and Inoue]{KulaSANER2014}
Raula~Gaikovina Kula, Daniel~M. German, Takashi Ishio, and Katsuro Inoue.
\newblock Trusting a library: A study of the latency to adopt the latest maven
  release.
\newblock In \emph{22nd IEEE International Conference on Software Analysis,
  Evolution, and Reengineering, SANER 2015, Montreal, Canada, March 2-6, 2015},
  2015.

\bibitem[Lehman(1996)]{Lehman:1996}
M.~M. Lehman.
\newblock Laws of software evolution revisited.
\newblock In \emph{Proceedings of the 5th European Workshop on Software Process
  Technology}, EWSPT '96, pages 108--124, London, UK, UK, 1996.
  Springer-Verlag.
\newblock ISBN 3-540-61771-X.

\bibitem[Lungu(2008)]{lungu2008}
Mircea Lungu.
\newblock Towards reverse engineering software ecosystems.
\newblock In \emph{Intl. Conf. on Soft. Maint. and Evo. (ICSME)}, 2008.

\bibitem[McDonnell et~al.(2013)McDonnell, Ray, and Kim]{McDonnell2013}
Tyler McDonnell, Baishakhi Ray, and Miryung Kim.
\newblock {An empirical study of API stability and adoption in the android
  ecosystem}.
\newblock \emph{IEEE International Conference on Software Maintenance, ICSM},
  pages 70--79, 2013.
\newblock ISSN 1063-6773.
\newblock \doi{10.1109/ICSM.2013.18}.

\bibitem[Mens et~al.(2014)Mens, Claes, and Grosjean]{Mens:ECOS}
Tom Mens, Maelick Claes, and Philippe Grosjean.
\newblock Ecos: Ecological studies of open source software ecosystems.
\newblock In \emph{Soft. Main. Reeng. and Rev. Eng. (CSMR-WCRE)}, pages
  403--406, Feb 2014.

\bibitem[Mileva et~al.(2009)Mileva, Dallmeier, Burger, and Zeller]{Mileva:2009}
Yana~Momchilova Mileva, Valentin Dallmeier, Martin Burger, and Andreas Zeller.
\newblock Mining trends of library usage.
\newblock In \emph{Proc. Intl and ERCIM Principles of Soft. Evol. (IWPSE) and
  Soft. Evol. (Evol) Workshops}, IWPSE-Evol '09, pages 57--62, New York, NY,
  USA, 2009. ACM.

\bibitem[Plate and Ponta(2015)]{Plate:ICSME2015}
Henrik Plate and Sabetta~Antonino Ponta, Serena~Elisa.
\newblock Impact assessment for vulnerabilities in open-source software
  libraries.
\newblock In \emph{Proceedings of the 31st International Conference on Software
  Maintenance and Evolution}, ICSME '15, Breman, Germany, 2015. IEEE Computer
  Society.

\bibitem[Raemaekers et~al.(2012)Raemaekers, van Deursen, and
  Visser]{RaemaekersICSM}
S.~Raemaekers, A.~van Deursen, and J.~Visser.
\newblock Measuring software library stability through historical version
  analysis.
\newblock In \emph{Proc. of Intl. Comf. Soft. Main. (ICSM)}, pages 378--387,
  Sept 2012.

\bibitem[Raemaekers et~al.(2014)Raemaekers, van Deursen, and
  Visser]{Raemaekers2014}
S.~Raemaekers, A.~van Deursen, and J.~Visser.
\newblock Semantic versioning versus breaking changes: A study of the maven
  repository.
\newblock In \emph{Source Code Analysis and Manipulation (SCAM), 2014 IEEE 14th
  International Working Conference on}, pages 215--224, Sept 2014.

\bibitem[Robbes et~al.(2012)Robbes, Lungu, and R\"{o}thlisberger]{Robbes:2012}
Romain Robbes, Mircea Lungu, and David R\"{o}thlisberger.
\newblock How do developers react to api deprecation?: The case of a smalltalk
  ecosystem.
\newblock In \emph{Proceedings of the ACM SIGSOFT 20th International Symposium
  on the Foundations of Software Engineering}, FSE '12, pages 56:1--56:11, New
  York, NY, USA, 2012. ACM.
\newblock ISBN 978-1-4503-1614-9.

\bibitem[Rogers(2003)]{DoI}
Everett~M. Rogers.
\newblock \emph{Diffusion of innovations}.
\newblock Free Press, NY, 5th edition, 08 2003.
\newblock ISBN 0-7432-2209-1, 978-0-7432-2209-9.

\bibitem[Sawant et~al.(2016)Sawant, Robbes, and Bacchelli]{Sawant2016}
Anand~Ashok Sawant, Romain Robbes, and Alberto Bacchelli.
\newblock On the reaction to deprecation of 25,357 clients of 4+1 popular java
  apis.
\newblock In \emph{Proceedings of the 32th IEEE International Conference on
  Software Maintenance and Evolution.}, 2016.

\bibitem[Sch\"{a}fer et~al.(2008)Sch\"{a}fer, Jonas, and Mezini]{Schafer:2008}
Thorsten Sch\"{a}fer, Jan Jonas, and Mira Mezini.
\newblock Mining framework usage changes from instantiation code.
\newblock In \emph{Proceedings of the 30th International Conference on Software
  Engineering}, ICSE '08, pages 471--480, New York, NY, USA, 2008. ACM.
\newblock ISBN 978-1-60558-079-1.

\bibitem[Teyton et~al.(2014)Teyton, Falleri, Palyart, and Blanc]{Teyton:2014}
Cédric Teyton, Jean-Rémy Falleri, Marc Palyart, and Xavier Blanc.
\newblock A study of library migrations in java.
\newblock \emph{Journal of Software: Evolution and Process}, 26\penalty0 (11),
  2014.

\bibitem[Wittern et~al.(2016)Wittern, Suter, and Rajagopalan]{Wittern:MSR2016}
Erik Wittern, Philippe Suter, and Shriram Rajagopalan.
\newblock A look at the dynamics of the javascript package ecosystem.
\newblock In \emph{Proc. of Work. Conf. on Mining Soft. Repo. (MSR2016)}, 2016.

\bibitem[Wu et~al.(2015{\natexlab{a}})Wu, Khomh, Adams, Gu{\'{e}}h{\'{e}}neuc,
  and Antoniol]{Wu2015APIUsages}
Wei Wu, Foutse Khomh, Bram Adams, Yann-Ga{\"{e}}l Gu{\'{e}}h{\'{e}}neuc, and
  Giuliano Antoniol.
\newblock {An exploratory study of api changes and usages based on apache and
  eclipse ecosystems}.
\newblock \emph{Empirical Software Engineering}, pages 1--47,
  2015{\natexlab{a}}.
\newblock ISSN 1573-7616.
\newblock \doi{10.1007/s10664-015-9411-7}.

\bibitem[Wu et~al.(2015{\natexlab{b}})Wu, Serveaux, Gu{\'e}h{\'e}neuc, and
  Antoniol]{Wu:2015APIEvoluion}
Wei Wu, Adrien Serveaux, Yann-Ga\"{e}l Gu{\'e}h{\'e}neuc, and Giuliano
  Antoniol.
\newblock The impact of imperfect change rules on framework api evolution
  identification: An empirical study.
\newblock \emph{Empirical Softw. Engg.}, 20\penalty0 (4):\penalty0 1126--1158,
  August 2015{\natexlab{b}}.
\newblock ISSN 1382-3256.
\newblock \doi{10.1007/s10664-014-9317-9}.

\bibitem[Xia et~al.(2013)Xia, Matsushita, Yoshida, and Inoue]{Xia2013}
Pei Xia, Makoto Matsushita, Norihiro Yoshida, and Katsuro Inoue.
\newblock Studying reuse of out-dated third-party code in open source projects.
\newblock In \emph{Japan society for Software Science and Technology}, Computer
  Software, Vol.30, No.4, pp.98-104, 2013.

\bibitem[Xing and Stroulia(2007)]{Xing2007}
Zhenchang Xing and Eleni Stroulia.
\newblock {API-evolution support with diff-catchup}.
\newblock \emph{Software Engineering, IEEE Transactions on}, 33:\penalty0
  818--836, 2007.
\newblock \doi{10.1109/TSE.2007.70747}.

\end{thebibliography}

\end{document}